\documentclass[useAMS,usenatbib]{mn2e}

\usepackage{amssymb}
\usepackage{aas_macros}
\usepackage{natbib}
\usepackage{times}
\usepackage{graphicx}
\newcommand{\yr}                         {\,{\rm yr}}

\newcommand{\Gyr}                      {\,{\rm Gyr}}

\newcommand{\kpc}                      {\,{\rm kpc}}
\newcommand{\Mpc}                      {\,{\rm Mpc}}
\newcommand{\Msun}                    {\,{\rm M}_\odot}

\newcommand{\hkpc}                     {\,h^{-1}\,{\rm kpc}}
\newcommand{\hMpc}                    {\,h^{-1}\,{\rm Mpc}}
\newcommand{\hMsun}                  {\,h^{-1}\,{\rm M}_\odot}
\newcommand{\kms}                      {\,\,{\rm km}\,\,{\rm s}^{-1}}
\newcommand{\cmcubed}              {\,\,{\rm cm}^{-3}}
\newcommand{\Zsun}                     {\,{\rm Z}_\odot}

\newcommand{\ergs}                     {\,{\rm erg}\,\,{\rm s}^{-1}}
\newcommand{\ergscm}                 {\,{\rm erg}\,{\rm s}^{-1}\,{\rm cm}^{-2}}
\newcommand{\K}                          {\,{\rm K}}
\newcommand{\keV}                      {\,{\rm keV}}
\newcommand{\nospacehMpc}      {h^{-1}\,{\rm Mpc}}
\newcommand{\LKsun}                  {\,{\rm L}_{{\rm K},\odot}}

\newcommand{\gimic}        {\textsc{gimic}}
\newcommand{\owls}          {\textsc{owls}}
\newcommand{\gadget}      {\textsc{Gadget3}}
\newcommand{\subfind}     {\textsc{Subfind}}
\newcommand{\cloudy}      {\textsc{Cloudy}}

\newcommand{\galexev}         {\textsc{Galaxev}}
\newcommand{\apec}              {\textsc{Apec}}

\newcommand{\einstein}             {\textit{Einstein}}
\newcommand{\rosat}             {\textit{ROSAT}}
\newcommand{\asca}             {\textit{ASCA}}
\newcommand{\chandra}         {\textit{Chandra}}
\newcommand{\xmm}             {\textit{XMM-Newton}}

\newcommand{\twomass}       {\textsc{2Mass}}

\newcommand{\lstar}             {$L_\star$}

\newcommand{\sfr}                {$\dot{M}_\star$}
\newcommand{\lx}                 {$L_{\rm X}$}
\newcommand{\lk}                 {$L_{\rm K}$}

\newcommand{\mstar}           {$M_{\star}$}

\newcommand{\mhalo}          {$M_{200}$}
\newcommand{\rhalo}           {$r_{200}$}

\newcommand{\ZFe}               {$Z_{\rm Fe}$}
\newcommand{\ZFesun}       {${\rm Z}_{\rm Fe,\odot}$}

\def\lsim{\mathrel{\lower0.6ex\hbox{$\buildrel {\textstyle <}
 \over {\scriptstyle \sim}$}}}

\title[Enriching the hot circumgalactic medium]{Enriching the hot circumgalactic medium}
 \author[R.~A.~Crain et al.]  {\parbox[h]{160mm}{
    Robert A. Crain$^{1}$\thanks{E-mail: crain@strw.leidenuniv.nl}
    Ian G. McCarthy$^{2,3}$,
    Joop Schaye$^{1}$,
    Tom Theuns$^{4,5}$ and\\
    Carlos S. Frenk$^{4}$ 
}
  \vspace{6pt}\\
  $^1$Leiden Observatory, Leiden University, PO Box 9513, 2300 RA Leiden, Netherlands \\
  $^2$Astrophysics and Space Research Group, School of Physics and Astronomy, University of Birmingham, Edgbaston, Birmingham B15 2TT\\ 
  $^3$Astrophysics Research Institute, Liverpool John Moores University, Birkenhead, CH41 1LD\\ 
  $^4$Institute for Computational Cosmology, Department of Physics, University of Durham, South Road, Durham, DH1 3LE\\
  $^5$Department of Physics, University of Antwerp, Campus Groenenborger, Groenenborgerlaan 171, B-2020 Antwerp,  Belgium}

\begin{document}

\date{\today}
\pagerange{\pageref{firstpage}--\pageref{lastpage}} \pubyear{2011}

\maketitle

\label{firstpage}

\begin{abstract}
  Simple models of galaxy formation in a cold dark matter universe
  predict that massive galaxies are surrounded by a hot,
  quasi-hydrostatic circumgalactic corona of slowly cooling gas, predominantly
  accreted from the intergalactic medium (IGM). This prediction is
  borne out by the recent cosmological hydrodynamical simulations of
  Crain et al., which reproduce observed scaling relations between the
  X-ray and optical properties of nearby disc galaxies. Such coronae
  are metal poor, but observations of the X-ray emitting circumgalactic medium (CGM) of local galaxies typically indicate enrichment to near-solar iron abundance, potentially signalling a shortcoming in current models of galaxy formation. We show here that, while the hot CGM
  of galaxies formed in the simulations is typically metal poor in a
  mass-weighted sense, its X-ray luminosity-weighted metallicity is
  often close to solar. This bias arises because the soft X-ray
  emissivity of a typical $\sim 0.1\keV$ corona is dominated by collisionally-excited metal ions that are
  synthesised in stars and recycled into the hot CGM. We find that
  these metals are ejected primarily by stars that form in-situ to the
  main progenitor of the galaxy, rather than in satellites or external
  galaxies. The enrichment of the hot CGM therefore proceeds in an
  `inside-out' fashion throughout the assembly of the galaxy: metals are transported from the central galaxy by supernova-driven
  winds and convection over several gigayears, establishing a strong negative radial metallicity gradient. Whilst metal ions synthesised by stars are necessary to produce
  the X-ray emissivity that enables the hot CGM of isolated galaxies
  to be detected with current instrumentation, the electrons that collisionally excite them are equally important. Since our simulations indicate that the electron density of hot coronae is dominated by the metal-poor gas accreted from the
  IGM, we infer that the hot CGM observed via X-ray emission is the outcome of both hierarchical accretion and stellar recycling. 
\end{abstract}



\section{Introduction}
\label{sec:introduction}

Circumgalactic coronae in quasi-hydrostatic equilibrium around massive
galaxies, formed by the accretion and shock-heating of intergalactic
gas, are a simple but fundamental prediction of galaxy formation
models in a cold dark matter universe
\citep[][]{White_and_Frenk_91,Kauffmann_White_and_Guiderdoni_93,Cole_et_al_00,Hatton_et_al_03,Bower_et_al_06,Croton_et_al_06,Monaco_Fontanot_and_Taffoni_07,Somerville_et_al_08,Benson_10}.
Their properties were first explored analytically by \citet[hereafter
WF91]{White_and_Frenk_91}, who posited that coronae associated with
\lstar\ galaxies at $z=0$ have temperatures of $T\sim10^6\K$, and emit
$\sim 10^{41}-10^{43}\ergs$ of cooling radiation, primarily in the
form of soft ($\sim 0.1\keV$) X-ray photons. If present in the local
Universe, circumgalactic reservoirs of this luminosity should have
been detected, if not by previous generations of X-ray satellites, 
certainly by \chandra\ and \xmm.

The well-documented failure of the \einstein\ and \rosat\ X-ray
observatories to detect hot atmospheres associated with massive disc
galaxies
\citep[e.g.][]{Bregman_and_Glassgold_82,Vogler_Pietsch_and_Kahabka_95,Bregman_and_Houck_97,Fabbiano_and_Juda_97,Benson_et_al_00}
has often been regarded as a fundamental challenge to galaxy formation
theory. More recently, the greater sensitivity of the \chandra\ and
\xmm\ observatories has led to detections of a hot circumgalactic
medium (CGM) associated with both isolated disc galaxies
\citep{Strickland_et_al_04,Wang_05,Tullmann_et_al_06,Li_Wang_and_Hameed_07,Jeltema_Binder_and_Mulchaey_08,Owen_and_Warwick_09,Rasmussen_et_al_09,Sun_et_al_09,Li_et_al_11}
and isolated elliptical galaxies, with no obvious signature of a
contribution from active galactic nuclei
\citep[AGN,][]{David_et_al_06,Mulchaey_and_Jeltema_10}. However, these
data have not established the validity of the canonical galaxy
formation framework because the inferred X-ray luminosity is typically
one to two orders of magnitude fainter than predicted by analytic
models. Moreover, the origin of the hot gas in these examples remains
controversial.

Most commonly, extra-planar X-ray emission associated with disc
galaxies is interpreted as a signature of hot ejecta from Type II
supernovae (SNe) entrained in galaxy-wide outflows.  \textit{Prima
  facie}, this interpretation is not without merit since several
studies report a correlation between the extra-planar X-ray luminosity
(\lx) and star formation indicators such as H$\alpha$ line emission or
broad-band optical/IR luminosity
\citep[e.g.][]{Strickland_et_al_04,Rasmussen_Stevens_and_Ponman_04,Tullmann_et_al_06,Sun_et_al_07,Li_and_Wang_12}.
Particularly spectacular cases such as the archetypal local starburst
system, M82, bolster this interpretation because of the biconical
morphology of their X-ray surface brightness contours and the presence
of coincident optical line emission
\citep[][]{Strickland_et_al_04,Veilleux_Cecil_and_Bland-Hawthorn_05}.

In contrast to disc galaxies, the association of hot circumgalactic
gas with elliptical galaxies has been established for many years
\citep[e.g.][]{Forman_et_al_79,Forman_Jones_and_Tucker_85,Fabbiano_89}.
In common with disc galaxies, however, the hot gas is commonly
interpreted as having an internal origin\footnote{Since we are
  mainly concerned here with $\sim$\lstar\ galaxies, we do not consider
  galaxies with significant AGN activity, as were considered
  theoretically by e.g.
  \citet{Ciotti_and_Ostriker_97,Ciotti_and_Ostriker_07}, nor those
  confined by the potential of galaxy groups.} as the ejecta
associated with evolved stellar populations such as Type Ia SNe and
asymptotic giant branch (AGB) stars \citep[see
e.g.][]{Mathews_90,Ciotti_et_al_91,Mathews_and_Brighenti_03,Parriott_and_Bregman_08}.
This interpretation seems reasonable because evolved stellar
populations unquestionably return a significant mass of gas to the
interstellar medium (ISM) when integrated over the lifetime of the
galaxy.  Moreover, a broad correlation between \lx\ and the stellar
mass of the galaxy (\mstar, or proxies such as \lk) has now been
firmly established in deep \chandra\ observations after careful
removal of contaminating X-ray sources
\citep{Boroson_Kim_and_Fabbiano_11}.

The interpretation of diffuse X-ray emission as a signature of
internal, stellar processes, rather than an accreted coronal reservoir,
therefore presents a challenge to current models of galaxy formation.
Motivated by this challenge, we conducted a study \citep[][hereafter
Paper I]{Crain_et_al_10} using the cosmological hydrodynamic
simulations of the \textsc{Galaxies-Intergalactic Medium Interaction
  Calculation} \citep[\gimic,][hereafter C09]{Crain_et_al_09_short} to
investigate the origin and nature of the hot CGM associated with an
unprecedentedly large sample of $\sim500$ \lstar\ simulated galaxies. Our analysis demonstrated that M82-like systems, whose X-ray luminosity is dominated by hot gas entrained in outflows, are indeed found in the \gimic\ simulations. However, just as in the local Universe, such systems are rare. The majority of $\sim$\lstar\ galaxies instead develop extended, quasi-hydrostatic coronae, primarily through the shock heating and adiabatic compression of gas accreted
from the intergalactic medium (IGM), supplemented by relatively small
amounts of gas recycled through the galaxy by stellar evolution
processes.  

This picture appears similar to that posited by the analytic model of WF91 but, crucially, the X-ray luminosity of the hot CGM of \gimic\ galaxies is one to two orders of magnitude fainter (at fixed halo mass) than predicted by WF91. The more detailed and accurate treatment of gas accretion and feedback in the \gimic\ simulations highlights that the discrepancy stems primarily from inaccurate assumptions regarding the physical structure of the corona adopted by WF91. The X-ray luminosity of \gimic\ galaxies is therefore broadly consistent with measurements inferred from \chandra\ and \xmm\ observations of local galaxies. For example,  \citet{Anderson_and_Bregman_11} and \citet{Dai_et_al_12} recently reported detections of luminous coronae associated with the giant disc galaxies NGC1961 and UGC 12591, respectively, whose inferred coronal luminosities correspond closely with those of similarly massive disc galaxies in \gimic.

It was noted in Paper I that \gimic\ galaxies exhibit a tight
correlation between \lx\ and halo mass (\mhalo), and this naturally
gives rise to a broad correlation between \lx\ and both stellar mass
(\mstar) and the star formation rate (\sfr) since, to first order,
both of these quantities correlate with \mhalo. Hence, these observed
correlations do not necessarily imply an internal origin for the hot
gas. This raises the appealing possibility of unifying the
interpretation of X-ray observations of disc and elliptical galaxies
in a simple model in which, as is inferred in all
cosmologically-motivated models of galaxy formation, the CGM of
galaxies is established by a mixture of intergalactic gas accreted
throughout their assembly history and small amounts of internally
recycled gas.

We intend to explore such a unification in detail elsewhere, and focus
here on what has long been perceived as a problem for the canonical
galaxy formation framework: that the hot CGM of $\sim$\lstar\ elliptical galaxies
appears to be enriched to a significant fraction of the solar
metallicity\footnote{The first metallicity measurement of
  the hot CGM of a \textit{late-type} galaxy (NGC891) was
  reported by \citet{Hodges-Kluck_and_Bregman_12} during the
  preparation of this manuscript. They report $Z\sim 0.1\Zsun$,
  consistent with the gas being primarily sourced by accretion from
  the IGM.}. Such high metallicities were first revealed by high spectral resolution X-ray spectroscopy of NGC 4636
\citep{Xu_et_al_02_short} and NGC 5044 \citep{Tamura_et_al_03} using 
\xmm's Reflection Grating Spectrometer (RGS). These measurements have
been broadly corroborated by imaging spectroscopy with \chandra\
\citep[][hereafter HB06; Ath07]{Humphrey_and_Buote_06,Athey_07},
which, at the cost of lower spectral resolution, offers a greater
effective area, wider bandwidth, better isolation of the diffuse gas,
and the possibility of spatially excising contaminating flux from
resolved X-ray point sources.  At first sight, the presence of
enriched hot gas in ellipticals supports a simple `internal origin'
for the gas, whereby its metallicity directly reflects that of the
evolved stellar populations comprising the galaxy \citep[which are
approximately near-solar,
e.g.][]{Trager_et_al_00a,Trager_et_al_00b,Terlevich_and_Forbes_02,Gallazzi_et_al_06},
and appears difficult to reconcile with the notion that the hot CGM
contains a large amount of metal-poor gas accreted from the IGM.

In this paper, we extend the analysis of the hot CGM of $z=0$
\gimic\ galaxies presented in Paper I, to confront the simulations
with the observational constraints on coronal metallicities presently
available in the literature. We demonstrate that the simulations are
compatible with the data, and elucidate the nature of the enrichment
of the hot CGM of $\ga$ \lstar\ galaxies, demonstrating how, where and when
the metals were injected. Building on the arguments presented in Paper
I, we consider this compatibility to be further evidence that X-ray
data support the canonical view of galaxy formation in the CDM
paradigm, in a fashion that is orthogonal to, for example, the
confrontation of galaxy formation models with optical/infra-red data
(e.g. the stellar mass function or the Tully -Fisher relation). This
paper is laid out as follows. In \S~\ref{sec:methods} we provide a
brief overview of the \gimic\ simulations and our analysis of them. We
confront these analyses with X-ray data in
\S~\ref{sec:coronal_metallicities}, and demonstrate how the enrichment
proceeds in \S~\ref{sec:dissecting_enrichment}. Finally, we discuss and
summarise our findings in \S~\ref{sec:discussion}.

\section{Methods}
\label{sec:methods}

\subsection{Simulations}
\label{sec:simulations}

The \gimic\ simulations are described in detail in C09, where thorough discussions of the generation of the initial conditions, the simulation algorithms, and initial results may be found. Techniques for computing gas-phase X-ray luminosities and stellar optical luminosities are described in detail in Paper I. Here, we present only a brief overview and limit the description to aspects that are specifically relevant to this study, referring the reader to C09 and Paper I for additional details.

Using `zoomed' initial conditions, \gimic\ follows with full gas
dynamics and, at relatively high resolution for cosmological volumes,
the evolution of five roughly spherical regions of comoving radius
$r\sim20\hMpc$, drawn from the Millennium Simulation
\citep{Springel_et_al_05_short}. The remainder of the
$500^3\,(\nospacehMpc)^3$ Millennium Simulation volume is modelled
with collisionless particles at much lower resolution, thus correctly
following the cosmic large-scale structure. In order to trace a wide
range of cosmic environments, the regions were chosen such that their
overdensities deviate by $(-2$, $-1$, $0$, $+1$,$+2)\sigma$ from the
cosmic mean, where $\sigma$ is the rms mass fluctuation on the
$20\hMpc$ radial scale of the regions, at $z=1.5$.

The \gimic\ initial conditions were generated at intermediate
resolution\footnote{We reserve the term `low resolution' for the
  original realisation of the Millennium Simulation.} ($m_{\rm gas} =
1.16\times 10^7\hMsun$) and high resolution ($m_{\rm gas} = 1.45\times
10^6\hMsun$), and ran with gravitational force softening for the
high-resolution particles of $\epsilon_{\rm phys}^{\rm max}
= (1.0,0.5)\hkpc$ for the intermediate- and high-resolution runs 
respectively. The relatively large cosmological volume traced with gas
dynamics produces a sample of over 600 relatively isolated $\sim$\lstar\
galaxies, making \gimic\ complementary to ultra-high resolution
simulations ($m_{\rm gas} \sim 10^4\hMsun$) of individual galaxies.

The cosmological parameters adopted for \gimic\ are the same as those
assumed by the Millennium Simulation and correspond to a $\Lambda$CDM
cosmogony with $\Omega_{\rm m} = 0.25$, $\Omega_\Lambda = 0.75,
\Omega_{\rm b} = 0.045$, $\sigma_8 = 0.9$, $H_0=100\,h\,{\rm km s^{-1}
  Mpc^{-1}}$, $h=0.73$, $n_{\rm s} = 1$ (where $n_{\rm s}$ is the
spectral index of the primordial power spectrum). The value of
$\sigma_8$ is approximately $2\sigma$ higher than inferred from the
most recent cosmic microwave background (CMB) and large-scale
structure data \citep{Komatsu_et_al_11_short}, which will affect the
abundance of \lstar\ galaxies somewhat, but should not significantly
affect their individual properties.

The initial conditions were evolved from $z=127$ to $z=0$ with the TreePM-SPH code \gadget, a substantial upgrade of \textsc{Gadget2} \citep{Springel_05} that features improved load-balancing \citep[as in][]{Springel_et_al_08} and incorporates a wide array of new baryon physics modules. Radiative cooling rates are computed on an element-by-element basis by interpolating pre-computed tables, generated with \cloudy\ \citep[version 07.02,][]{Ferland_et_al_98}, that specify cooling rates as a function of density, temperature and redshift, and that account for the presence of the CMB and a spatially-uniform \citet{Haardt_and_Madau_01} ionising UV/X-ray background \citep[for further details, see][]{Wiersma_Schaye_and_Smith_09}. The ionising background is imposed at $z=9$ and effectively quenches star formation in small haloes \citep[][C09]{Okamoto_Gao_and_Theuns_08}. Star formation is implemented following the prescription of \citet{Schaye_and_Dalla_Vecchia_08}, which reproduces the observed Kennicutt-Schmidt law \citep{Kennicutt_review_98} by construction. 

Feedback from SNe is implemented using the kinetic wind model of
\citet{Dalla_Vecchia_and_Schaye_08}, with an initial wind velocity
$v_{\rm w} = 600\kms$ and mass-loading parameter (the ratio of mass of
gas that receives an impulsive kick to the mass of stars formed) of
$\beta=4$. This scheme is energetically feasible, requiring
approximately 80 percent of the total energy available from Type II
SNe for our adopted choice of initial mass function (IMF), that due to
\citet{Chabrier_03}. This choice of parameters produces a good match to
the peak of the cosmic star formation history \citep[see
C09,][]{Schaye_et_al_10}. The timed release of individual elements by
massive stars (Type II SNe and asymptotic giant branch stars) and
intermediate-mass stars (Type Ia SNe) is incorporated following the
prescription of \citet{Wiersma_et_al_09}. A set of 11 elements (H, He, C, Ca, Fe, Mg, N, Ne, O, S, Si), representing the most important
radiative coolants, are followed, enabling us to determine the element
abundances of individual SPH and star particles.

The \gimic\ simulations are particularly well suited to the study of
$\sim$\lstar\ galaxies. As shown in Paper I, the implementation of
efficient (but energetically feasible) feedback from SNe largely
prevents overcooling on the mass scale of \lstar\ galaxies, and is key
to the reproduction of the observed X-ray scaling relation presented
in that study. Indeed, \gimic\ accurately reproduces the rotation speeds and star formation efficiencies of $z=0$ disc galaxies for $10^9 \lesssim M_\star < 10^{10.5} \Msun$, although galaxies with $M_\star \ga 10^{11} \Msun$ do still suffer from some overcooling
\citep{McCarthy_et_al_12b}. Moreover, \citet{Font_et_al_11} demonstrated that \lstar\ galaxies in \gimic\ exhibit satellite luminosity functions and stellar spheroid surface brightness distributions that are comparable to those of the Milky Way and M31, whilst \citet{McCarthy_et_al_12a} further demonstrated that this
correspondence extends also to their global structure and kinematics.

\subsection{Galaxy \& halo identification}
\label{sec:galaxy_identification}

We identify bound haloes using the \subfind\ algorithm presented by
\citet{Dolag_et_al_09}, which extends the standard implementation
\citep{Springel_et_al_01} by including baryonic particles when
identifying self-bound substructures. This procedure first finds dark
matter haloes using a friends-of-friends (FoF; \cite{DEFW85})
algorithm, with the standard linking length in units of the
interparticle separation ($b=0.2$) It also associates all baryonic
particles with their nearest neighbour dark matter particle. The
aggregated properties of baryonic particles associated with grouped
dark matter particles define the baryonic properties of each FoF
group. Halo substructures are then identified using a topological
unbinding algorithm. This provides an unambiguous definition of a
\textit{galaxy} within the simulations, namely the set of star
particles bound to individual subhaloes. The gas bound to each subhalo
then forms the ISM and CGM; these phases are delineated by the density
threshold for the onset of star formation, $n_{\rm H} = 0.1\cmcubed$.
Since individual haloes may have more than one `subhalo', they may
host more than one galaxy. In analogy to semi-analytic models, the
stars associated with the most massive subhalo of a FoF group are
defined as central galaxies, whilst stars associated with
substructures are satellites. When referring to the mass of haloes, we
adopt spherical overdensity \citep[SO,][]{Lacey_and_Cole_94} masses,
centred on the most bound particle within FoF groups, with radius such
that the mean enclosed density is 200 times the critical density, i.e.
\mhalo.

\subsection{Simulated galaxy sample}
\label{sec:sample}

As in Paper I, we draw galaxies from all five intermediate-resolution
\gimic\ regions, and use the high-resolution $-2\sigma$ region as a
benchmark for brief convergence tests, which are presented in
Appendix \ref{sec:numerical_convergence}. We impose a number of selection criteria in order to
construct a sample similar to observational samples: we select central
(i.e. the most massive galaxy within an FoF halo) galaxies whose
stellar mass lies in the range $10^{10} < M_\star < 10^{11.5} \Msun$,
corresponding to a fairly broad mass range bracketing \lstar. We
additionally require that the subhalo with which each galaxy is
associated should account for at least 90 per cent of the mass of the parent
FoF group, i.e. $M_{\rm sub} > 0.9M_{\rm FoF}$. This criterion is
intended to select only isolated systems by excluding those that are
interacting or are members of galaxy groups or clusters. These
selection criteria result in a sample of 617 galaxies at $z=0$.

As shown in \citet{McCarthy_et_al_12b}, the
intermediate-resolution \gimic\ simulations accurately reproduce the
stellar mass - rotation speed (or `Tully-Fisher') relation at $z=0$
for galaxies with $10^9 \lesssim M_\star < 10^{10.5}
\Msun$. These galaxies are consistent with the \mstar-\mhalo\ relation
inferred from stacked
weak lensing + satellite kinematics analyses \citep[see][and
references therein]{Dutton_et_al_10}. \citet{McCarthy_et_al_12b} also found that the most
massive galaxies in our sample ($M_\star \ga 10^{11} \Msun$, corresponding to $K$-band luminosities
$L_{\rm K} \gtrsim 10^{11} L_{{\rm K},\odot}$) exhibit some
overcooling, since our feedback scheme becomes inefficient within
massive haloes (see C09). We retain these galaxies in our sample in
order to connect with observations of similarly massive ellipticals,
but caution that these galaxies are likely hosted by dark matter
haloes that are insufficiently massive, and exhibit star formation
rates at low-redshift that are too high. 

\subsection{Computing optical \& X-ray luminosities}
\label{sec:computing_luminosities}

Optical and X-ray luminosities are computed in postprocessing. The
former are calculated individually for star particles, considering
them as simple stellar populations (SSP). Star particles are formed
with the IMF due to  \citet{Chabrier_03}, and their spectral energy
distribution (SED) is derived by interpolation over the \galexev\
models of \citet{BC03}. The optical luminosity of each particle is
obtained by integrating the product of its SED with the appropriate
filter transmission function; the overall broadband luminosity of a
galaxy is then defined as the sum of the luminosities of all star
particles comprising a galaxy. Since we are primarily concerned here
with $K$-band luminosities, we do not expect dust extinction to be an
important consideration.  

\begin{table}
\begin{center}
\caption{Abundances of the 11 elements tracked self-consistently within the \gimic\ simulations, and used to compute the radiative cooling rate and X-ray luminosity of gas, presented for the \citet[][]{Anders_and_Grevesse_89} abundance standard (assumed by the \apec\ model), and that adopted by the \gimic\ simulations, which is derived from \cloudy\ (version 07.02). We follow the convention $\log_{10}N_{\rm H}=12$.}
\begin{tabular}{ l c c }
\hline
\hline
Element & AG89 & \cloudy\ \\
\hline
Hydrogen  & $12.00$ & $12.00$ \\
Helium    & $10.99$ & $11.00$ \\
Carbon    & $8.56$  & $8.39$ \\
Calcium   & $6.36$  & $6.36$ \\
Iron      & $7.67$  & $7.45$ \\
Magnesium & $7.58$  & $7.54$ \\
Nitrogen  & $8.05$  & $7.93$ \\
Neon      & $8.09$  & $8.00$ \\
Oxygen    & $8.93$  & $8.69$ \\
Sulphur   & $7.21$  & $7.26$ \\
Silicon   & $7.55$  & $7.54$ \\
\hline
\label{tab:abundances}
\end{tabular}
\end{center}
\end{table}

Gas phase X-ray luminosities are also computed on a per-particle
basis, following:
\begin{equation}
L_{\rm X}  = n_{{\rm e}} n_{{\rm i}} \Lambda V,
\end{equation}
where $n_{\rm e}$ and $n_{\rm i}$ are the number densities of
electrons and ions respectively, $\Lambda$ is the cooling
function in units of $\ergscm$ (integrated over an appropriate
passband, for example 0.5-2.0\keV), and $V = m_{\rm gas} / \rho$,
where $m_{\rm gas}$ is the gas particle mass and $\rho$ the gas
particle density. We compute $\Lambda$ by interpolating a pre-computed
table generated using the Astrophysical Plasma Emission Code
\citep[\apec, v1.3.1, see][]{Smith_et_al_01} under the assumption that
the gas is an optically thin plasma in collisional ionisation
equilibrium. \apec\ cooling rates are computed on an
element-by-element basis for each particle,
\begin{equation}
\Lambda_j(T_j) = \frac{Z_{j,k}}{Z_{k, \rm AG89}}\sum_{k=1}^{N}\lambda_{k}(T_j),
\label{eq:emissivity}
\end{equation}
\noindent where $T_j$ is the gas temperature of the $j^{\rm th}$ particle, $Z_{j,k}$ is the metal mass fraction of the element $k$ for this particle, and $\lambda_{k}(T_j)$ is the cooling function for the element $k$. The total cooling rate is arrived at by summing over the $N=11$ most important elements for cooling (H, He, C, Ca, Fe, Mg, N, Ne, O, S, Si), which are individually and self-consistently 
tracked during the simulation. \apec\ assumes the solar abundance ratios of \citet[][hereafter AG89]{Anders_and_Grevesse_89}, but it is straightforward to rescale the cooling rates for arbitrary abundances, which we do here on a particle-by-particle basis, normalising by the metal mass fractions assumed by AG89 ($Z_{k, \rm AG89}$), to achieve consistency with our chemodynamics implementation. For completeness, we present the solar abundances adopted by AG89 and \gimic\ in Table~\ref{tab:abundances}. The latter corresponds to the abundances assumed by \cloudy\ \citep[version 07.02, last described by][]{Ferland_et_al_98}. Our scheme excludes gas within the ISM \citep[particles with non-zero star formation rate, see][]{Schaye_and_Dalla_Vecchia_08} since we assign these particles a temperature of $T=10^4{\rm K}$.

\subsection{Computing iron abundances}
\label{sec:computing_iron_abundances}

Since iron has the strongest diagnostic lines in the soft X-ray band, \ZFe\ is the most commonly used proxy for gas-phase metallicities in low temperature plasmas \citep{Loewenstein_and_Mushotsky_98}. We therefore adopt the mass-weighted iron mass fraction as a metallicity proxy, computed thus: 
\begin{equation}
Z_{\rm Fe}^{\rm mw} = \frac{\sum_j^N{Z_{{\rm Fe},j}m_j}}{\sum_j^N{m_j}},
\label{eq:ZFe_mw}
\end{equation}
\noindent where $Z_{{\rm Fe},j}$ is the SPH kernel-weighted iron abundance of the $j^{\rm th}$ gas particle (see discussion in \S~\ref{sec:metal_mixing}), and $m_j$ is its mass. The sum runs over the $N$ gas particles associated with each galaxy that have a non-zero soft X-ray luminosity (and thus comprise the hot CGM, see \S~\ref{sec:computing_luminosities}). 

In Paper I, we cautioned that the sensitivity of the X-ray emissivity
of gas to density, temperature and metallicity can lead to an
unrepresentative view of the overall state of hot gas. In more
specific terms, the unavoidable luminosity-weighting of X-ray
observations can yield measurements that differ markedly from an
ideal, mass-weighted measurement. The most direct comparison between
simulated and observed coronae is obtained by deriving \lx-weighted
quantities in the former. Therefore, we also define a
luminosity-weighted metallicity indicator, that is appropriate when
comparing with observationally-inferred metallicities, by exchanging
the particle masses ($m_j$) in Eqn.~\ref{eq:ZFe_mw} for their SPH kernel weighted soft
X-ray luminosities ($L_{{\rm X},j}$):
\begin{equation}
Z_{\rm Fe}^{\rm lw} = \frac{\sum_j^N{Z_{{\rm Fe},j}L_{{\rm X},j}}}{\sum_j^N{L_{{\rm X},j}}}.
\label{eq:ZFe_lw}
\end{equation}
In general. we use the mass weighted measure to assess the physical state of the hot CGM, and reserve the luminosity weighted measure for direct comparsions with observationally-inferred measurements.

\subsection{Metal mixing}
\label{sec:metal_mixing}

The cooling rate, and therefore star formation rate, of cosmic gas is
strongly affected by its metallicity, so the mixing of stellar
nucleosynthetic products into the ISM and CGM is clearly of significance for models of galaxy evolution. The
process of metal mixing, however, remains poorly described by
theory, and poorly constrained by observation. 

Nonetheless, a common theme amongst both types of study is that mixing is not particularly efficient, across a broad
range of scales. On relatively small scales (a few parsecs), the Eulerian simulations of \citet{Parriott_and_Bregman_08}, and particularly
\citet{Bregman_and_Parriott_09}, indictate that much of the ejecta
from evolved stars and planetary nebulae remain unmixed with the
hot ISM/CGM. This picture appears to be corroborated by observations
of the structure of stellar ejecta at both infra-red
\citep{Wareing_et_al_06} and ultra-violet
\citep{Martin_et_al_07_short} wavelengths. On significantly larger
scales, \citet{Schaye_Carswell_and_Kim_07} argue that the large population of
compact, intergalactic CIV absorbers seen in their sample of
QSO-sightlines is evidence for inefficient mixing of metals entrained
in winds and transported from the ISM into the CGM/IGM.

The physical interpretation of the distribution and transport of
metals seen in our simulations therefore requires an understanding of
the behaviour and limitations of the numerical implementation of these
processes. The
formulation of SPH naturally defines three spatial scales for which a
discussion of metal mixing is appropriate: scales smaller than, comparable to, or larger than the SPH kernel (i.e. the smoothing length). 

On the smallest scales, the nucleosynthetic products of stellar evolution are implicitly assumed to be completely mixed, since the metal mass released by the SSP represented by each star particle is assigned to neighboring SPH particles. Clearly, each particle is the smallest unit over which mass can be distributed, so it is therefore possible that on subgrid scales mixing is overestimated. Such scales are, by definition, below the resolution limit of the simulation, and as such physical insight should not be drawn from the simulations on these scales.

On scales comparable to the kernel, the finite numerical sampling of
the fluid, and the absence of implicit diffusion of scalar quantities
associated with SPH particles, suppresses the mixing of entropy and
metals\footnote{It is not
  clear that Eulerian algorithms present a more appealing alternative,
  since they can in fact mix gas too efficiently
  \citep[e.g.][]{Mitchell_et_al_09}.} \citep{Wadsley_Veeravalli_and_Couchman_09}. This shortcoming is clearly of
relevance here, and it has been shown that the metallicity of
circumgalactic gas in simulations is increased when subgrid models for
efficient turbulent mixing are imposed
\citep{Shen_Wadsley_and_Stinson_10}. However, we opt against imposing
such a scheme here, since it represents represent an attempt to
resolve a numerical problem with a model of a poorly understood physical process. 

We intend to explore small scale diffusion in future work, using SPH
schemes augmented with higher-order dissipation switches
\citep[e.g.][]{Read_and_Hayfield_12} that diffuse scalar quantities
implicitly. Here, we follow \citet{Wiersma_et_al_09} and minimise the
sampling problem by computing cooling rates, X-ray luminosities and overall coronal iron mass
fractions (Eqns.~\ref{eq:ZFe_mw} \& \ref{eq:ZFe_lw}) using
kernel-weighted element abundances averaged over a given SPH
particle's neighbours. The resulting smoothed metallicity complements
the `particle metallicity' that is the sum of the element mass
fractions carried by each particle. This process has the potential to
bias the recovered metallicity of hot circumgalactic particles,
because smoothed metallicities are computed using all SPH neighbour
particles, irrespective of their thermal state. The smoothed element
abundances of hot particles close to the disc-corona interface are
therefore computed from both cold (i.e. ISM) and hot (coronal)
particles. We gauge the degree to which mixing on the scale of the
kernel influences our conclusions, by comparing results derived from
particle and smoothed metallicities. We conclude from this analysis,
presented in Appendix \ref{sec:metallicity_smoothing}, that mixing on
the scale of the kernel has no bearing on our findings.

On larger scales still, simulations indicate that the distribution of
metals is governed chiefly by winds driven by energetic feedback
\citep{Wiersma_Schaye_and_Theuns_11}, a process self-consistently
modelled by our simulations. Whilst the standard implementation of SPH
is well-understood to suppress the development of Rayleigh-Taylor and
Kelvin-Helmholtz instabilities that would assist the mixing of metals
entrained in winds into their ambient medium, the inefficient mixing
inferred by \citet{Schaye_Carswell_and_Kim_07} on these relatively
large scales suggests that it is reasonable to conclude that
predictions of the large-scale distribution of metals derived from our
simulations are robust to the implementation of metal mixing adopted by our simulations. It is the metal distribution on these large
scales that is the central the focus of this study.

\subsection{Stacked maps}
\label{sec:stacks}

\begin{figure*}
\includegraphics[width=\textwidth]{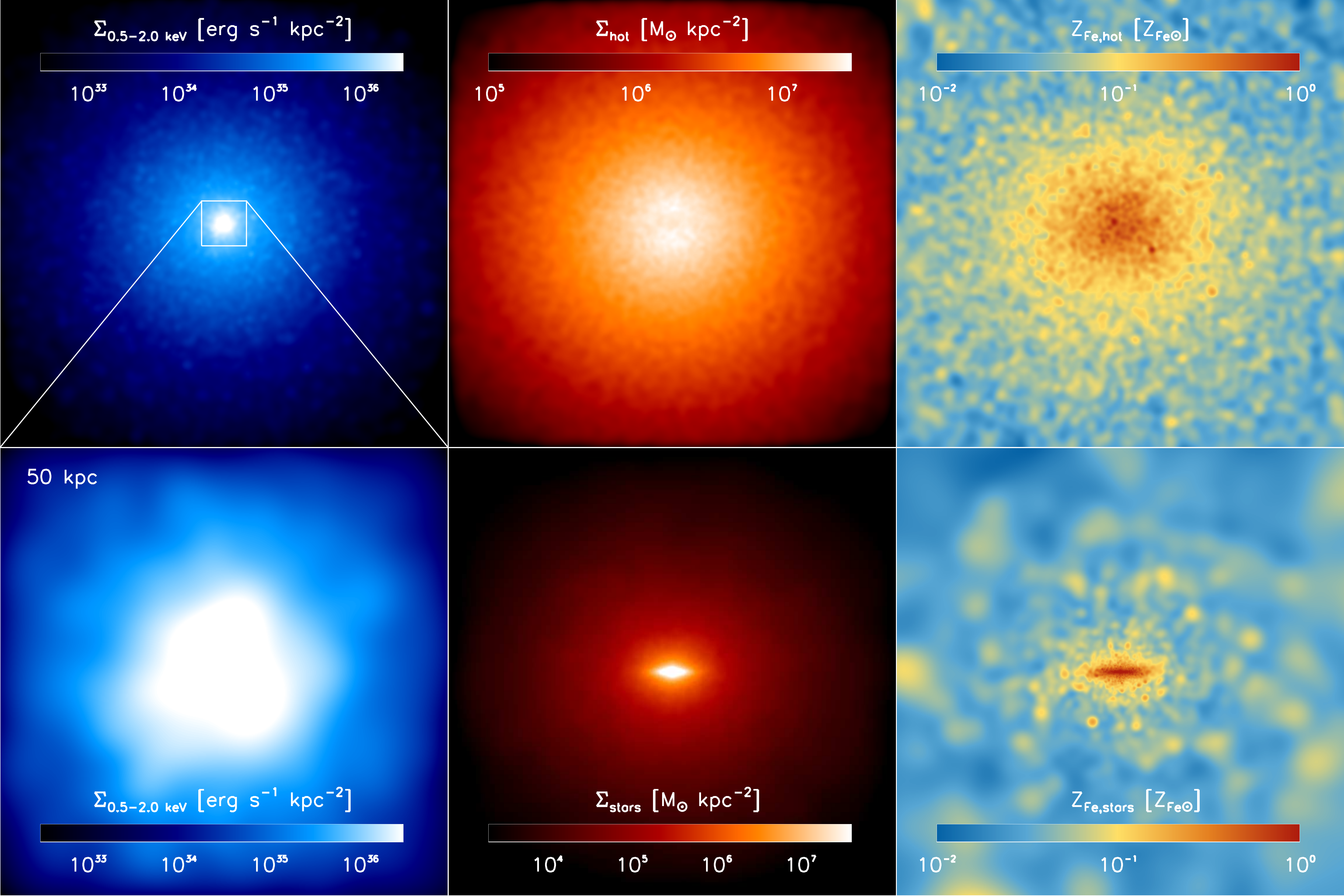}
\caption{Stacked maps of the baryons associated with the 148 \gimic\ galaxies (from our sample of 617) with halo mass in the range $10^{12.25} < M_{200}/{\rm M}_\odot < 10^{12.50}$. The galaxies have been rotated so that the angular momentum of stars is parallel to the vertical axis prior to stacking. The maps are 500\kpc\ across, with the exception of the bottom left panel, which is zoomed to 50\kpc. \textit{Left column}: the soft X-ray (0.5-2.0\keV) surface brightness. \textit{Centre column}: the surface density of hot gas (\textit{top}) and stars (\textit{bottom}). \textit{Right column}: the iron abundance of hot gas (\textit{top}) and stars (\textit{bottom}). The X-ray emission is markedly more centrally peaked than the mass distribution of the hot CGM, owing to its strong dependence on density, metallicity and temperature. A halo-wide iron abundance gradient is evident in the hot CGM, that does not trace the underlying distribution of stars, indicating the prevalence of metal transport.}
\label{fig:maps}
\end{figure*}

To illustrate the correspondence of the simulated galaxies with observable properties, we present in Fig.~\ref{fig:maps} maps derived from stacking the 148 galaxies from our sample whose halo masses lie in the range $10^{12.00} < M_{200} [{\rm M_\odot}] < 10^{12.25}$. The maps are 500\kpc\ across, with the exception of the bottom left panel, which is zoomed-in to 50\kpc, and in each case the galaxies have been rotated into the edge-on orientation by aligning the vertical axis with the angular momentum vector of all stars within 35\kpc\ of the galactic centre. The left hand column shows the soft X-ray surface brightness ($\Sigma_{\rm 0.5-2.0 keV}$) of the hot CGM, the central column shows the surface density of the hot CGM ($\Sigma_{\rm hot-CGM}$; \textit{top}) and of stars ($\Sigma_*$; \textit{bottom}), whilst the right hand column shows the iron abundance of the hot CGM ($Z_{\rm Fe, hot-CGM}$; \textit{top}) and of stars ($Z_{\rm Fe, *}$; \textit{bottom}). 

The alignment procedure is clearly effective: a disc can be seen in the stellar surface density map, as can the spheroidal stellar halo component that was explored by \citet{Font_et_al_11} and \citet{McCarthy_et_al_12a}. Similarly, the stellar iron abundance shows a clear disc plane, and a large-scale gradient similar to the metallicity gradients observed in nearby stellar haloes \citep{Foster_et_al_09,Spolaor_et_al_10}. The presence of strong disc features in these maps is unsurprising, as roughly two-thirds of the galaxies comprising our sample are disc-dominated (see Paper I). The hot gas maps exhibit circular surface density and soft X-ray surface brightness distributions, indicative of a quasi-spherical hot CGM associated with the galaxies. The coronal iron distribution is centrally peaked, but does not trace the underlying stellar iron abundance, indicating that metal transport is important for enriching the hot CGM. We return to the issue of metal transport in \S~\ref{sec:dissecting_enrichment}. 

\begin{figure}
\includegraphics[width=\columnwidth]{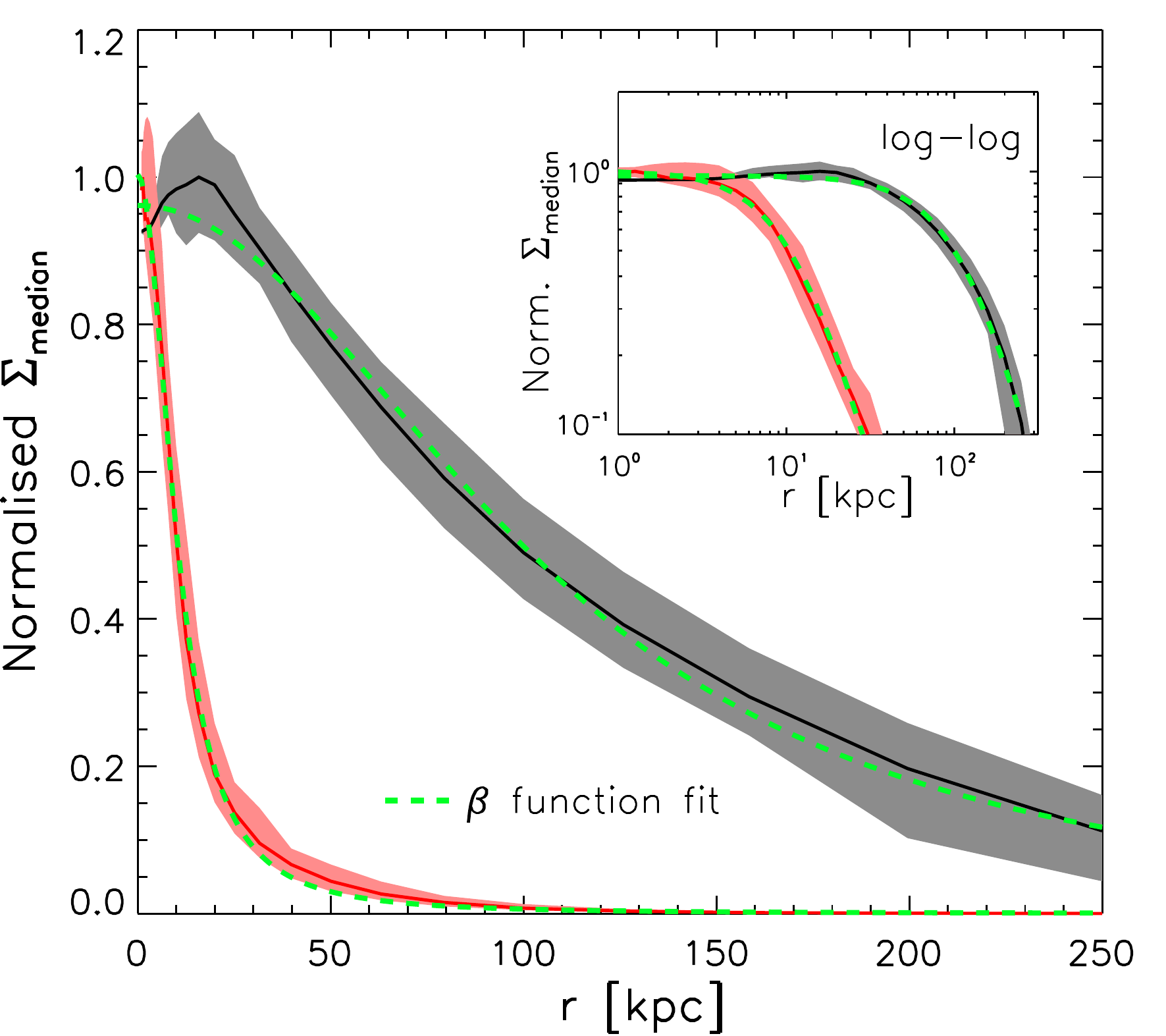}
\caption{Circularly-averaged median projected radial profiles of the
  surface density of the hot CGM (\textit{black}) and its soft X-ray
  surface brightness (\textit{red}), derived by radially binning the
  maps presented in Fig.~\ref{fig:maps}, and normalising to the
  maximum value of the median in each case. Shaded regions represent
  the scatter between $10^{\rm th}$ and $90^{\rm th}$ percentiles, and
  the inset plot presents the same data on logarithmic axes. Green
  dashed lines represent best fit $\beta$ functions to both profiles
  (see text for parameters). It is clear that the surface brightness is much more centrally peaked than the surface density, and therefore that X-ray emission does not trace the hot gas mass in a simple fashion.}
\label{fig:2d_profiles}
\end{figure}

A qualitative difference in the radial distribution of the coronal gas
and its X-ray luminosity is evident in the maps: the former is much
more extended than the latter. This follows from the complex mapping
of the state of the ionised plasmas ($\rho,T,Z$) into an X-ray
emissivity: bremsstrahlung radiation is weakly dependent on
temperature ($\propto T^{1/2}$) and strongly dependent on density
($\propto \rho^2$), whilst the metal line emission that dominates the emissivity of low temperature plasmas is also dependent on the abundance of metals. A quantitative assessment of the difference is
presented in Fig.~\ref{fig:2d_profiles}, where we bin the stacked maps
radially, and plot the radial surface density of the hot CGM
(\textit{black}) and its X-ray surface brightness
(\textit{red}). Solid lines show the median value of each quantity,
arbitrarily normalised to peak at a value of unity, whilst shaded
regions represent the scatter between the 10$^{\rm th}$ and 90$^{\rm
  th}$ percentiles. The inset plot presents the same data on
logarithmic axes, and the overplotted dashed green lines represent
fits, using the commonly adopted $\beta$-function:
\begin{equation}
\Sigma(r) \propto \left[1 + \left(\frac{r}{r_{\rm
        c}}\right)^2\right]^{(0.5-3\beta)},
\label{eq:beta_profile}
\end{equation}
to the stacked surface density and surface
brightness profiles, where $r_{\rm c}$ is the core
radius and $\beta$ characterises the profile shape. It is immediately evident that X-ray emission does
not trace the hot gas mass in a simple fashion: the surface brightness
profile is much more centrally peaked than the surface density
profile. This is reflected quantitatively in the least-squares reduced
parameters of the fitted $\beta$-functions: the surface density is
best described by ($r_{\rm c}, \beta = 120\kpc, 0.59)$, whilst the
surface brightness is best described by  ($r_{\rm c}, \beta =
12\kpc, 0.56)$. The recovered $\beta$ values are consistent with those
derived from a Monte Carlo Markov Chain analysis of stacked \rosat\ pointings of 2165 galaxies from the
\twomass\ Very Isolated Galaxy Catalogue, recently presented by
\citet{Anderson_Bregman_and_Dai_12}, and similar to the values
derived from deep observations of NGC 1961 and UGC 12591 by
\citet{Anderson_and_Bregman_11} and \citet{Dai_et_al_12},
respectively.  Those authors infer $1\kpc$ cores for NGC 1961 and UGC
12591, and constrain the stacked core radii to $0.5-5.0\kpc$. This is somewhat
smaller than the core radius we recover from our stacked surface
brightness profile, highlighting a potential shortcoming of the
simulations and an interesting avenue for future study. However,
having also fitted $\beta$-functions to the individual profiles of the
well-sampled galaxies in our sample, find that a wide range of core
radii are recovered; a small but significant fraction of the profiles ($\sim 15$
percent) are in fact best described by cusped profiles ($r_{\rm c} \rightarrow
0$). The simulations therefore yield a non-negligible fraction of
galaxies that are similar to those observed. Moreover, we caution that
the small sample of observationally inferred fits may be
unrepresentative of the population we explore with the simulations,
since galaxies with small core radius exhibit a greater peak surface brightness at
fixed luminosity, rendering them easier to detect \citep[c.f][]{Rasmussen_et_al_06}.

Besides highlighting that it is challenging to observe hot coronae at
radii significantly beyond the optical radius of the central galaxy,
Fig.~\ref{fig:2d_profiles} indicates that, in the absence of spatially-resolved spectroscopy (which is extremely challenging to realise for the hot CGM of galaxies with current instrumentation), the intrinsic luminosity weighting of observations biases the coronal metallicity inferred from X-ray spectroscopy towards the high surface brightness, metal rich gas close to galactic centres. The importance of this effect will be explored in the following section.

\section{The coronal metallicities of present-day galaxies}
\label{sec:coronal_metallicities}

In this section, we present the iron mass fractions of the hot CGM
associated with \gimic\ galaxies at $z=0$, and confront them with
observational measurements inferred from X-ray imaging spectroscopy.
We start with a brief discussion of observational studies in the
literature.

\subsection{Confrontation with observational data}
\label{sec:observational_data}

\begin{figure*}
\includegraphics[width=\columnwidth]{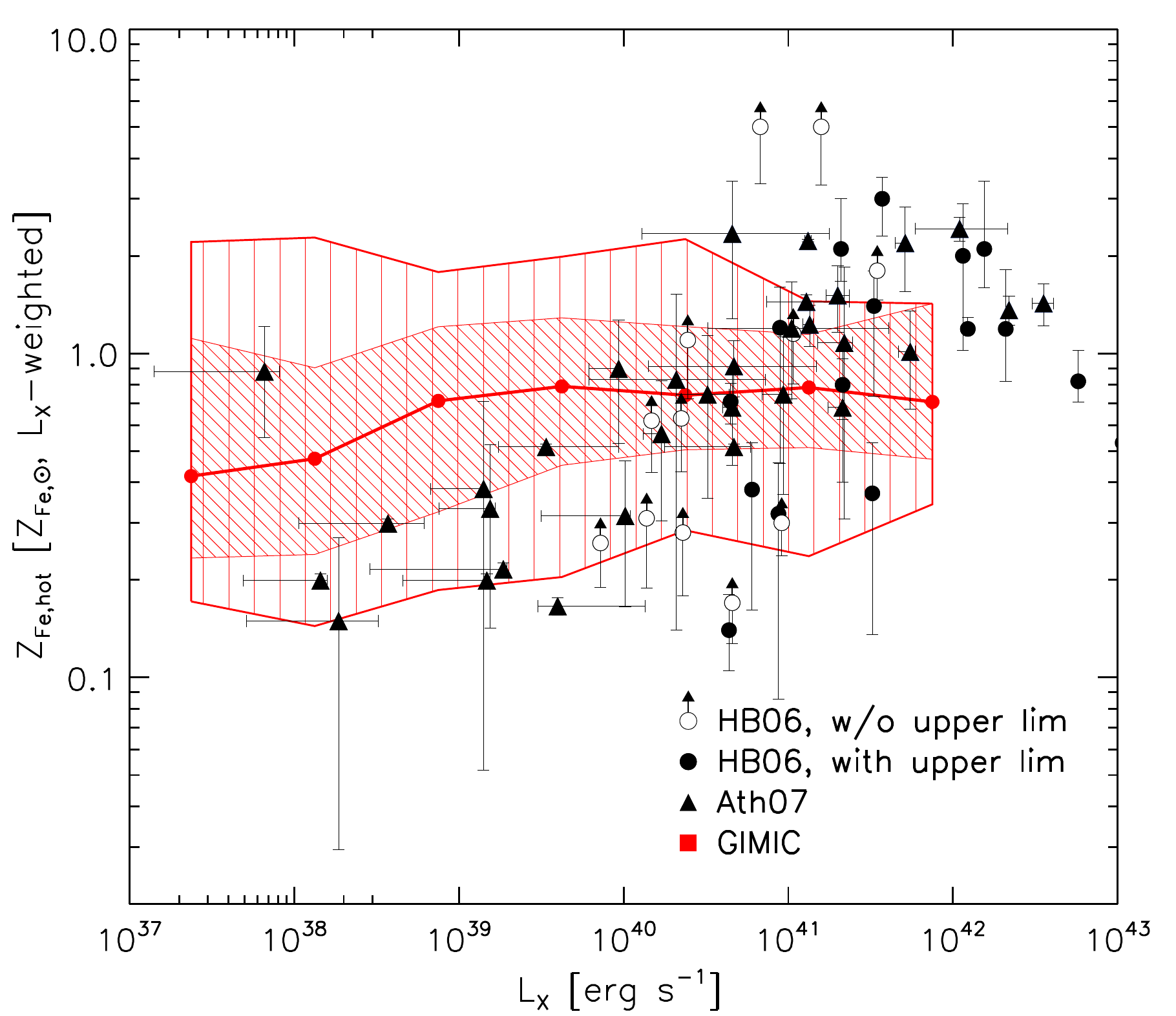}
\includegraphics[width=\columnwidth]{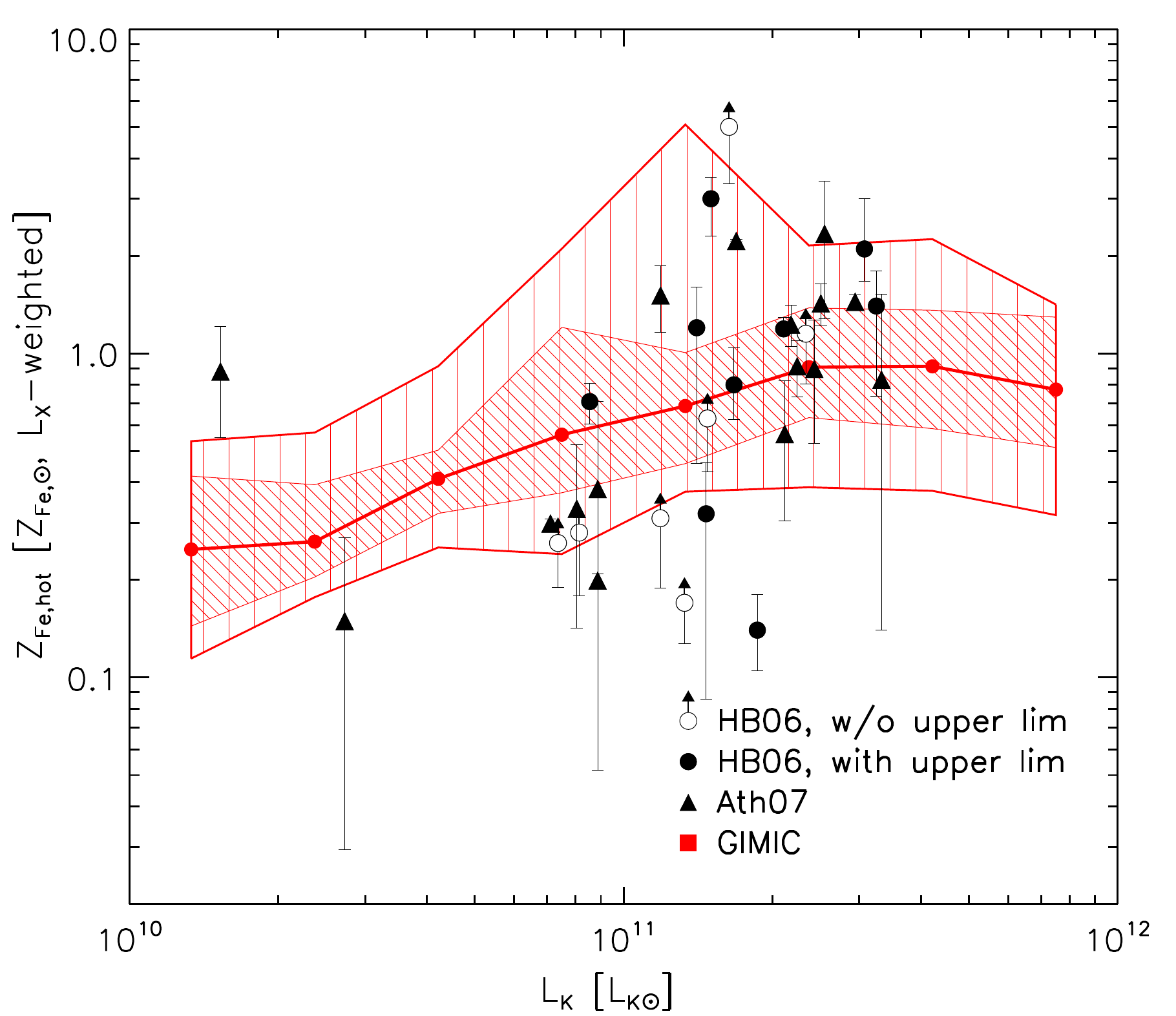}
\caption{The X-ray luminosity-weighted iron abundance of
  circumgalactic gas, as a function of soft X-ray luminosity
  (\textit{left}) and the $K$-band luminosity of galaxies
  (\textit{right}). The left-hand panel comprises the 28 nearby
  elliptical galaxies analysed by HB06 (\textit{circles}) and the 26
  (from a total sample of 54) nearby ellipticals from the sample of
  Ath07 (\textit{triangles}) that do not overlap with the HB06
  sample. The right hand panel comprises the subset of the overall
  galaxy sample for which $K$-band magnitudes are available in the
  2MASS Large Galaxy Atlas. Open circles with upward arrows represent
  cases where the HB06 spectral analysis yielded no formal upper limit
  on the iron abundance. Symbols are overplotted on the median
  \lx-weighted $Z_{\rm Fe}-L_{\rm X}$ and $Z_{\rm Fe}-L_{\rm K}$
  relations of galaxies in the \gimic\ simulations, with $1\sigma$ and
  $2\sigma$ scatter shown by the densely- and sparsely-hatched red
  regions respectively. Although the correlation between \ZFe\ and \lx,
  implied by the addition of the Ath07 sample to the HB06 sample, is
  not reproduced by the simulation (but see
  Appendix~\ref{sec:spectroscopy_comparison}), the
  normalisation of the observed metallicities is broadly reproduced by
  the simulations.} 
\label{fig:ZFe_Lx_Lk}
\end{figure*}

The excellent spatial resolution of \chandra\ has revolutionised
studies of metal enrichment in galaxies and galaxy groups. The
so-called `Fe-discrepancy' \citep{Arimoto_et_al_97,Renzini_97}, namely
the inference of highly subsolar iron abundances in early type
galaxies, has been shown to result from the inability of \rosat\ and
\asca\ to resolve coronal temperature gradients (HB06), that can lead
to underestimated abundances when isothermal spectral fits are adopted
\citep[e.g.][]{Buote_and_Canizares_94,Trinchieri_Kim_and_Fabbiano_94,Buote_and_Fabian_98}.
Early studies with \chandra\ (and \xmm) focussing on individual
systems indicated near-solar coronal iron mass abundances
\citep[e.g.][]{Buote_02,Gastaldello_and_Molendi_02,Buote_et_al_03b,O'Sullivan_et_al_03,Humphrey_Buote_and_Canizares_04,Kim_and_Fabbiano_04}.
Motivated to extend this discovery to a representative sample of
galaxies, HB06 uniformly reduced archival \chandra\ data for 28
nearby, early type galaxies with soft X-ray luminosities spanning the
range $10^{39}\lesssim L_{\rm X}~[\ergs] \lesssim 10^{43}$. They
report that their measurements support an absence of correlation
between \ZFe\ and \lx, and confirm that in all but one case (NGC
1553), the coronal iron abundance corresponds closely with that of the
stars. A similar study was presented by Ath07, who uniformly reduced
\chandra\ imaging spectroscopy data (from the ACIS-S3 instrument) for
54 nearby ($d \lesssim 110\Mpc$) early type galaxies. For the
brightest 22 galaxies in this sample, Ath07 inferred iron abundances
via spectral modelling adopting variable abundance ratios, and found
that only a narrow spread of ratios were required for good fits. The
lower signal-to-noise ratio of the remaining 32 fainter galaxies
required that their metallicities be derived subject to the assumption
that their abundance ratios follow those of the brighter galaxies. In
common with HB06, Ath07 report typical iron abundances that are
approximately solar. We note that 22 galaxies are present in both
samples.

In this study we adopt the HB06 sample of 28 galaxies, and supplement
it with the 32 galaxies from Ath07 that are not present in HB06; in
general these are fainter systems than were studied by HB06. For
reference, we show briefly in Appendix \ref{sec:spectroscopy_comparison} how closely the \ZFe\ and
\lx\ measurements inferred by these two studies match. Whilst we refer
the reader to HB06 and Ath07 for detailed descriptions of the data
reduction, it is appropriate to discuss key aspects of their analyses.
Firstly, Ath07 adopt the superceded solar abundance standard of
\citet{Anders_and_Grevesse_89}, whilst HB06 adopt the more recent
standard of \citet{Asplund_Grevesse_and_Sauval_05}, which is very similar to that adopted by \gimic. For consistency,
in what follows, we therefore multiply the iron abundances derived by Ath07 by a
factor of 1.65 to achieve consistency between the adopted standards. Where the signal-to-noise ratio of the X-ray spectroscopy allows, both studies fit plasma models with variable abundance ratios; in the event of
insufficient S/N, HB06 fix to solar ratios, whilst Ath07 adopt the
typical ratios of the high-S/N galaxies. HB06 report estimated `global
abundances', whereby systems with measureable abundance gradients are
fit by a power-law and extrapolated to 30 arcminutes. Ath07 report hot
gas abundances within one effective (optical) radius of the host
elliptical. Finally, we note that the spectral analysis performed by
HB06 produced no upper bound on the iron abundance of 12 of 28 galaxies
in their sample; we plot these are open circles with no upper error
bar.

We compare the \ZFe-\lx\ relation of the HB06 and Ath07 samples with
the \gimic\ results in the left-hand panel of
Fig.~\ref{fig:ZFe_Lx_Lk}. The observational measurements are overplotted on the median and $1\sigma$
and $2\sigma$ scatter (represented by the densely- and
sparsely-hatched regions, respectively) of the \lx-weighted coronal
iron abundance of the 617 galaxies in the \gimic\ $\sim$\lstar\ galaxy
sample. The \gimic\ sample does not extend to such high soft X-ray
luminosities ($> 10^{42}\ergs$) as the data, since overcooling in
haloes with these luminosities produces galaxies that are much more
massive than \lstar\ at $z=0$ and are hence excluded from our sample. 

The simulated galaxies exhibit no strong correlation between the \lx-weighted iron abundance and \lx, with a median \ZFe\ that is near-solar across five decades in \lx, and a $2\sigma$-scatter of approximately one decade in \ZFe. This absence of correlation suggests that simulations with efficient feedback in massive haloes would not generate significantly different iron abundances to
those shown here. In fact, \citet{McCarthy_et_al_10} showed that including efficient feedback in galaxy groups, implemented as AGN
feedback in the \owls\ simulations \citep{Schaye_et_al_10}, has almost no impact upon the iron injected into the hot intragroup/intracluster medium, even though the mass of stars in galaxy groups/clusters is supressed by a factor of four. This is because the additional iron that is synthesised in simulations without AGN feedback is mostly injected into the cold ISM (rather than the hot CGM), and is rapidly reincorporated into subsequent generations of stars.

HB06 concluded that their sample shows no
convincing evidence of a correlation between \ZFe\ and \lx. The
addition of the Ath07 data, which populates the \ZFe-\lx\ plane at
lower luminosities, indicates an increase in metallicity from low
to high luminosity systems. The simulated sample does not exhibit such
a pronounced correlation. Whilst this potentially represents a
shortcoming of the simulations, we note that the correlation remains ill-understood
observationally and, as discussed by
\citet{Buote_and_Fabian_98} and HB06, the recovery of \ZFe\ is
potentially hampered by complex, multi-component thermal structure in
the hot gas. This tends to bias the measurement towards low
abundances. A second potential shortcoming of the model is the absence
of galaxies with $Z_{\rm Fe} \gtrsim 2 \Zsun$, as are present in the
HB06 sample. We again caution, however, that the quoted uncertainties
potentially underestimate the true uncertainy of each measurement. As an example, we note that NGC~4365
features in both samples: HB06 infer a metallicity of $Z_{\rm Fe} = 5
\Zsun$, whilst Ath07 assigns it a sub-solar metallicity. This
uncertainty is discussed further in Appendix \ref{sec:spectroscopy_comparison}.

We stress that it is not our aim to advance the \gimic\ simulations as
a complete and detailed description of the CGM as observed in soft X-rays. Whilst a detailed dissection of
the establishment of the \ZFe\ - \lx\ relation is clearly an
interesting avenue for future study, it is our aim here to test whether observational
measurements rule out the general framework within which the hot CGM
is established primarily via the accretion and heating of metal-poor
intergalactic gas, as proposed in Paper I. The left-hand panel of
Fig.~\ref{fig:ZFe_Lx_Lk} demonstrates conclusively that the presence
of a hot CGM, dominated by low-metallicity gas accreted from the IGM, is
perfectly consistent with luminosity-weighted, near-solar
metallicities. In our cosmological simulations, in which most of the
hot CGM is of extragalactic origin, the median luminosity-weighted
iron abundance of the gas is typically $\sim0.4-0.7 \Zsun$, with
$2\sigma$ scatter between $\sim0.2-1.1\Zsun$, across five decades in
X-ray luminosity. Thus, our simulations demonstrate that the
often-used argument that high metallicities preclude a predominantly
extragalactic origin for the hot CGM of typical galaxies is not robust.

As a consistency check we plot, in the right-hand panel of
Fig.~\ref{fig:ZFe_Lx_Lk}, the \ZFe-\lk\ relation for the subset of
galaxies in the HB06 and Ath07 samples for which $K$-band magnitudes
are available in the 2MASS Large Galaxy Atlas
\citep{Jarrett_et_al_03}\footnote{Whilst all of the galaxies of the HB06 and Ath07 samples are included in the 2MASS Extended Source Catalogue, the integrated fluxes of these nearby galaxies are recorded inaccurately, because of their proximity to scan edges. The 2MASS Large Galaxy Atlas accounts for this inaccuracy by combining adjoining scans.}. The data exhibit a larger scatter
(approximately two decades in \ZFe\ at \lk$=10^{11}\LKsun$), which is
broadly consistent with the simulations. The simulations indicate a
weak correlation that is broadly consistent with the observational
measurements over this luminosity range. The
\lx-\lk\ relation for \gimic\ galaxies was presented in Paper I, where
it was shown to accurately reproduce the observed relation for disc
galaxies. We intend to present a comparison with both disc and
elliptical galaxies in a forthcoming paper (Crain et al. \textit{in
  prep}).

There is a reasonable match between the results of the simulated
galaxies and the observational measurements: nearly all of the
measurements lie within the $2\sigma$ scatter of the simulation
results, over five decades in \lx. This is remarkable given that the
parameters of the subgrid modules in the simulations were not tuned to
reproduce any particular X-ray scaling relations.  We conclude that
models of galaxy formation based upon the CDM paradigm in which the
hot CGM of $\ga$ \lstar\ galaxies is dominated by gas accreted from the IGM,
are consistent with the best available observational constraints on
the heavy element abundance of hot gas associated with such galaxies.

\subsection{Recycled mass and luminosity profiles}
\label{sec:rec_profiles}

As motivated in the Introduction, the correspondence of near-solar
coronal iron abundances measured in the data and in the simulations
may at first seem surprising, since it was shown in Paper I (see Fig.
12 of that study) that the hot CGM of the simulated galaxies is
dominated by metal-poor gas accreted from the IGM. To understand how
these findings are reconciled, we probe the enrichment of the
simulated coronae in greater detail in this section.

\begin{figure}
\includegraphics[width=\columnwidth]{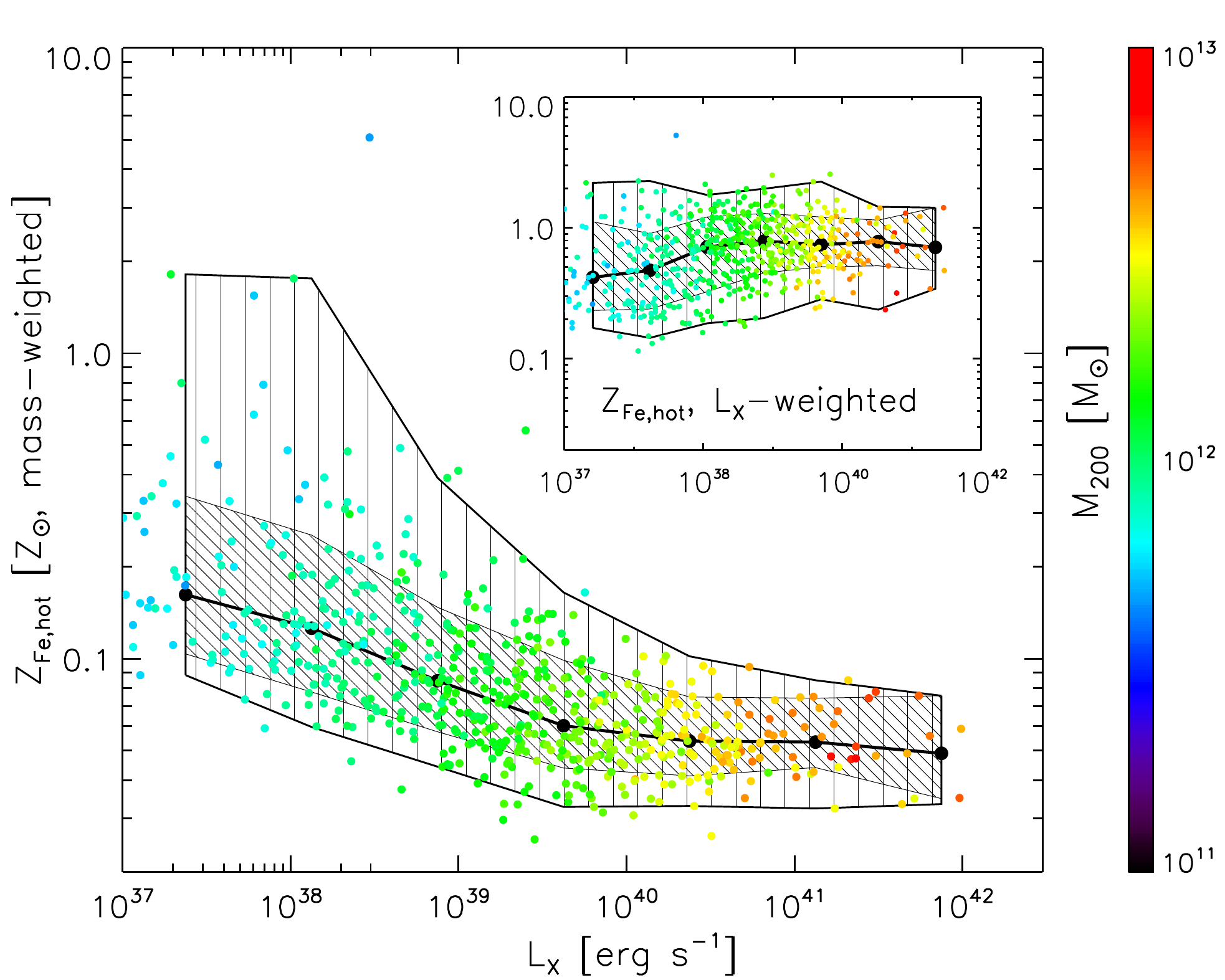}
\caption{The mass-weighted $Z_{\rm Fe}-L_{\rm X}$ relation of \gimic\ galaxies. Galaxies are represented by circles whose colour encodes the galaxy's halo mass, whilst the solid line, densely-hatched and sparsely-hatched regions show the median (hot gas) mass-weighted \ZFe, and its $1\sigma$ and $2\sigma$ scatter, respectively. For reference, luminosity-weighted measurements (as in the left-hand panel of Fig.~\ref{fig:ZFe_Lx_Lk}) are plotted in the inset. Although the \lx-weighted iron abundances are consistent with the near-solar values inferred from X-ray data, the hot CGM is metal-poor in a mass-weighted sense. Moreover, X-ray luminosity can be seen to correlate strongly with halo mass. These findings are expected if the hot CGM is established primarily via the accretion of gas from the IGM.} 
\label{fig:ZFe_Lx_mw}
\end{figure}

We remarked earlier, upon inspection of the surface density and surface brightness profiles presented in Fig.~\ref{fig:2d_profiles}, that luminosity-weighted X-ray measurements provide 
an unrepresentative view of the state of hot CGM, because the X-ray
emission does not trace the hot gas mass in a simple fashion. The
effect of this bias on the inferred metallicity of the hot CGM is
highlighted in Fig.~\ref{fig:ZFe_Lx_mw}, where we plot the
mass-weighted equivalent of Fig.~\ref{fig:ZFe_Lx_Lk} for the simulated
galaxy sample. Each galaxy is represented by a filled circle whose
colour encodes its halo mass, $M_{200}$, and a clear correlation between \lx\ and halo mass is evident (see also Fig.~6 of Paper I). Low luminosity galaxies
exhibit a slightly greater median iron abundance and a larger scatter
than their higher luminosity counterparts, because the hot CGM associated with these systems is comprised of a greater fraction of gas recycled from evolved stars than is the case for more massive galaxies, as we discuss below. However, the median value of \ZFe\ is typically $\lesssim 0.1\,\Zsun$ across the range of luminosities probed here. This contrasts with the \lx-weighted
measurement, which is near-solar across five decades in \lx. The
colour coding reveals the strong correlation between \lx\ and halo
mass (see also Fig.~6 of Paper I). For comparison, the inset panel
shows the luminosity-weighed measurements from which the median and
scatter shown in Fig.~\ref{fig:ZFe_Lx_Lk} were obtained.

\begin{figure*}
\includegraphics[width=\columnwidth]{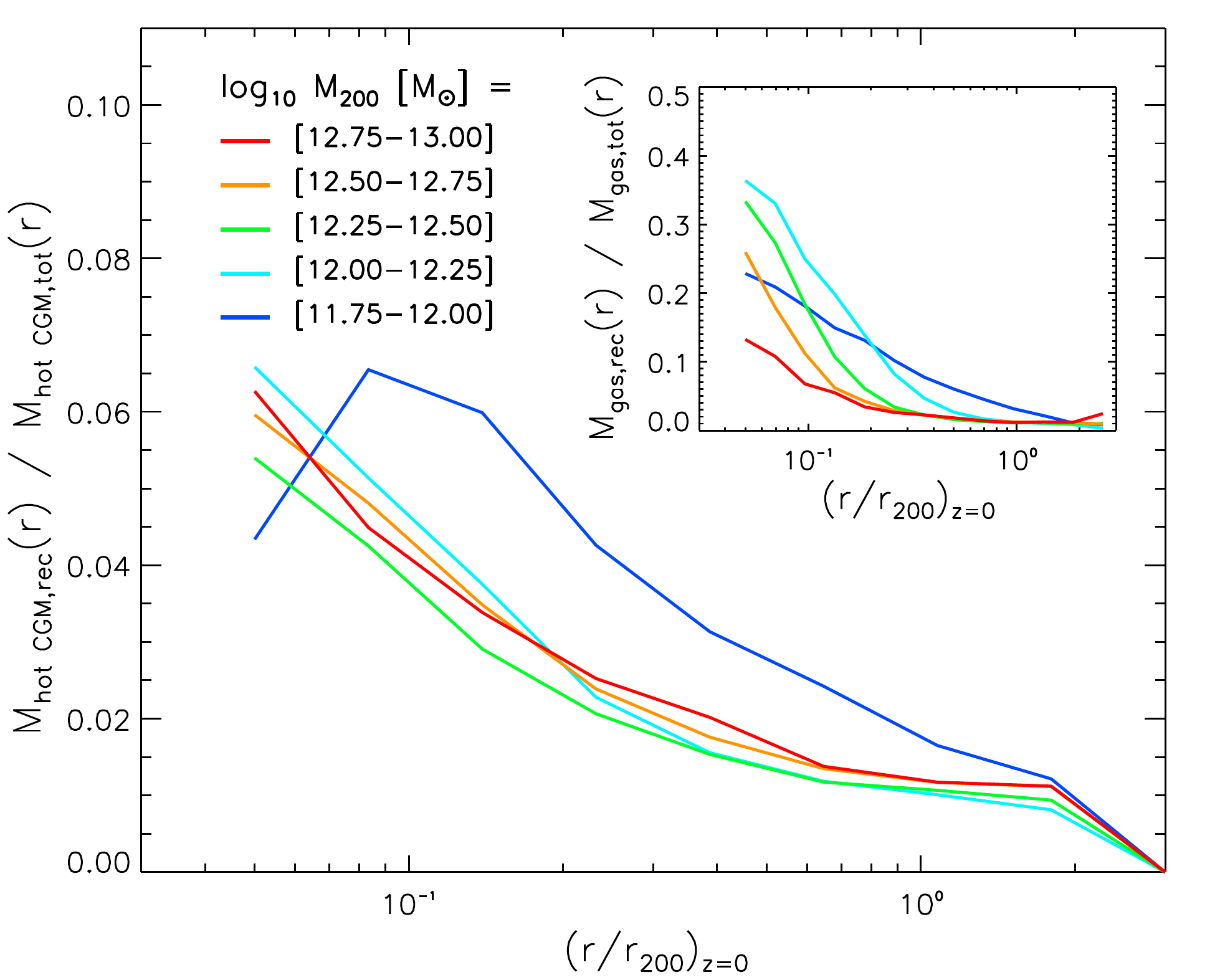}
\includegraphics[width=\columnwidth]{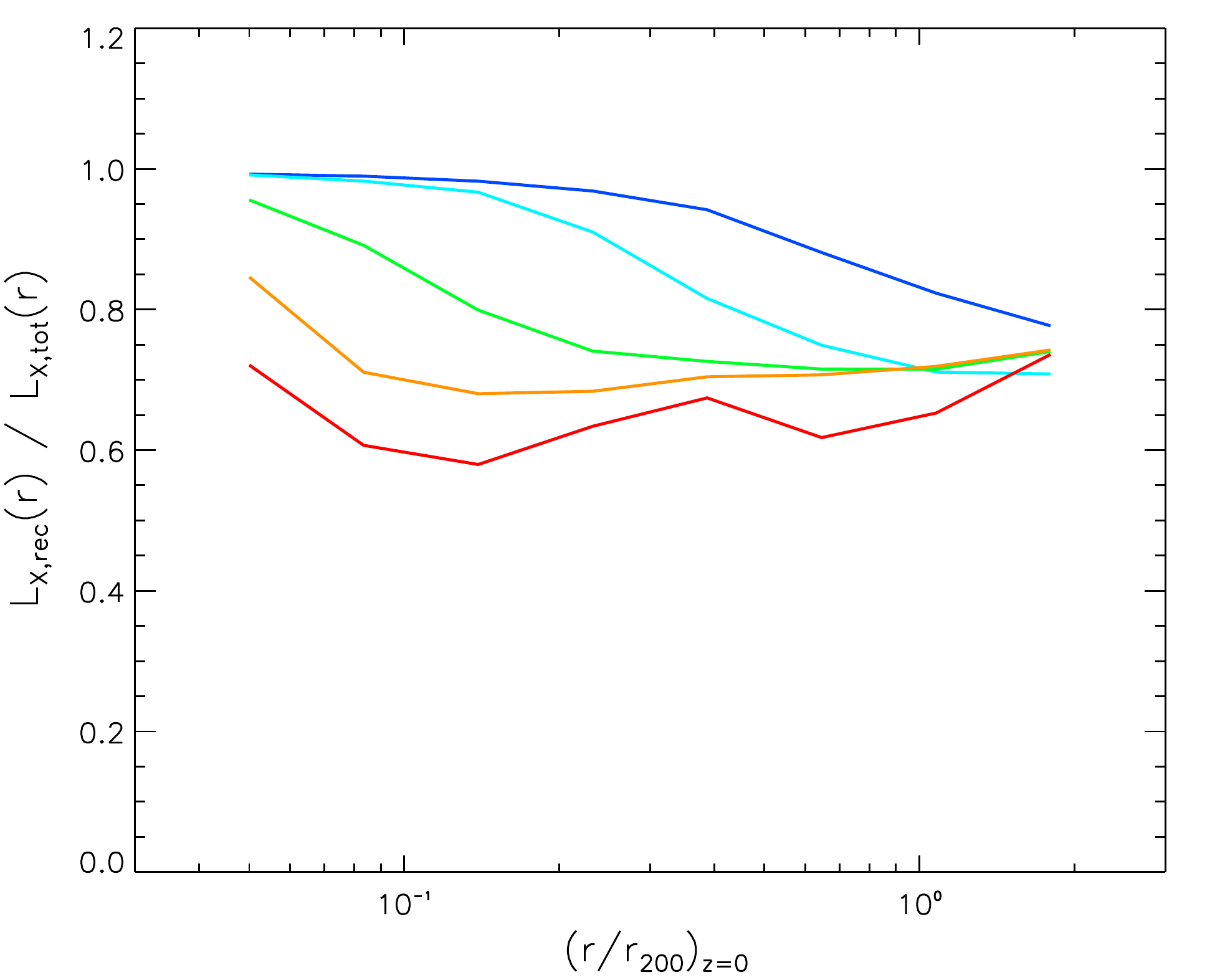}
\caption{\textit{Left:} Spherically-averaged radial profiles of the
  fraction, by mass, of the hot CGM contributed by gas recycled by
  stellar evolution. Galaxies have been binned by \mhalo\ (see
  legend). The inset shows the contribution to all halo gas (ISM +
  CGM), irrespective of thermal phase (note the different y-axis
  ranges). The contribution of recycled gas to the hot CGM is less
  than 10 percent at all radii (though the contribution to the ISM and
  cold CGM phases can be several times this). The small fraction of
  hot recycled gas is largely confined to small radii, close to the
  stars that produce it. \textit{Right:} As left, but showing the
  fraction of soft X-ray luminosity produced by recycled gas. In spite
  of recycled gas contributing a small fraction of the hot CGM mass,
  it dominates the diffuse soft X-ray emission of the galaxy at all
  radii. This leads to the marked difference between the mass-weighted
  and luminosity-weighted metallicity seen in Fig.~\ref{fig:ZFe_Lx_mw}. Note the markedly different scales on the y-axes of the three plots.}
\label{fig:radial_frec_profile}
\end{figure*}

The hot CGM of a typical $\sim$\lstar\ galaxy in our simulation sample
at $z=0$ is therefore metal poor, indicating that the hot gas is not
comprised primarily of the products of stellar evolution. It is possible
to check this explicitly, since the gas mass `recycled' by the stellar
populations represented by star particles is shared amongst their
neighboring SPH particles. Any difference between an SPH particle's
mass at $z=0$ and the start of the simulation is therefore due solely
to enrichment by stellar evolution. In
Fig.~\ref{fig:radial_frec_profile}, we separate the simulated sample
by halo mass into bins of $\Delta \log_{10} M_{200} [\Msun] = 0.25$.
The left-hand panel shows the spherically-averaged radial profile of
the median recycled mass fraction of particles comprising the hot CGM
and, for reference, we plot in the inset the equivalent profiles for
the entire CGM, irrespective of its thermal state. There is a strong
radial dependence, such that the products of stellar evolution are,
unsurprisingly, found close to the stars themselves, near the halo
centre (metals are, nonetheless, transported to large radius, as we
discuss in \S~\ref{sec:metal_transport}). The overall fraction of the
hot CGM contributed by recycling is greatest for the lowest mass
systems, as can also be seen in Fig.~\ref{fig:ZFe_Lx_mw}, for two
reasons. Firstly, the characteristic virial temperature of these
systems is low, so a smaller fraction of their accreted gas is X-ray
luminous (and hence becomes part of the `hot CGM'). Secondly, the baryon fraction of low mass haloes is lower than that of haloes hosting brighter galaxies because gas is more efficiently ejected by winds in the former\footnote{The baryon fraction of \gimic\ galaxies is a strong function of halo potential. At $z=0$, the typical baryon fraction increases smoothly from 10 percent to 90 percent of the cosmic fraction for galaxies with circular velocities between $100 \lesssim v_{\rm c} \lesssim 400\kms$ (see Fig.~12 of C09), and exhibits significant scatter. This value is particularly sensitive to the details of the adopted feedback prescription \citep[e.g.][]{Haas_et_al_12}, and it is likely that the baryon fraction of massive galaxies here is overestimated owing to the omission of efficient feedback in massive haloes, such as that from AGN.}. The metal mass injected into the hot CGM by stellar evolution is therefore diluted by a smaller mass of metal-poor gas accreted directly from the IGM.

The overall contribution of recycled gas to the hot CGM is, nonetheless, small at all radii for haloes of any mass: even near the halo centre, recycled gas comprises less than 10 percent of the
hot CGM (as also shown in Paper I), and integrated over the entire hot CGM
the contribution is typically only a few percent. By contrast, as
shown by the inset plot, recycled gas contributes a significant (but
not dominant) fraction when considering all gas bound to the halo
(i.e. ISM + all CGM gas, irrespective of thermal phase). Therefore a significant fraction of metals synthesised by the stars comprising today's \lstar\ galaxies have either been ejected into the IGM \citep[see also][]{Ciotti_et_al_91}, or recycled into a cool phase (i.e. the ISM or cold CGM) from which they can become locked up in subsequent generations of stars
\citep[e.g.][]{Leitner_and_Kravtsov_11}.

In the right-hand panel of Fig.~\ref{fig:radial_frec_profile} we show
the fraction of the diffuse soft X-ray luminosity produced by recycled
gas. We compute the contribution from recycled gas on a per-particle
basis, by equating the ratio $L_{\rm X,rec}/L_{\rm X,tot}$ to that of
the specific soft X-ray emissivities, $\Lambda_{\rm rec}/\Lambda_{\rm
  tot} = 1 - \Lambda_{\rm pri}/\Lambda_{\rm tot}$, where $\Lambda_{\rm
  pri}$ is the specific emissivity of the gas packet assuming it has primordial element abundances. As discussed by \citet{Wiersma_et_al_09}, this approach is accurate as long as the contribution of heavy elements to the free electron density is a small fraction of the total\footnote{For gas of solar composition, approximately one percent of electrons are contributed by metal ions.}. The emissivities are computed as per
Eq.~\ref{eq:emissivity}, summing over 11 elements for each particle,
albeit with the element abundances ($Z_k$ in Eq.~\ref{eq:emissivity})
of elements heavier than Helium fixed at zero in the case of
$\Lambda_{\rm pri}$. 

In contrast to the mass fraction shown in the left-hand panel, the
luminosity fraction due to recycled gas dominates at all radii. This is because the specific emissivity of a $\sim 0.1\keV$ plasma is extremely sensitive to the presence of metal ions. The effect is most dramatic in the case of low mass galaxies, which retain
a smaller fraction of their baryons in the form of an accreted, quasi-hydrostatic corona (see
C09), and in which galactic winds are more efficient at transporting
enriched gas from the galaxy into the CGM and IGM in the \gimic\
simulations. In addition, the bias between the recycled mass and luminosity fractions is expected to be largest for low mass galaxies, because the offset between the halo virial temperature, and the temperature at which the emissivity of collisionally-excited iron peaks, is greatest for these systems \citep[][]{vandeVoort_and_Schaye_13}. Nonetheless, the emissivity of the hot CGM associated
with even the most massive galaxies in our sample is dominated by
recycled gas. Since the hot CGM of typical galaxies remains extremely
challenging to detect in emission, we infer from our simulations that
the presence of metals in the hot CGM is a crucial aspect of its `observability', since they exhibit a far higher specific emissivity than hydrogen and helium ions.

It is important to clarify at this juncture what is meant by `recycled
gas'. In the context of Fig.~\ref{fig:radial_frec_profile}, it would be
more accurate to use the term `recycled ions': the mass of the
plasma is dominated by ions (as opposed to electrons), and when computing the plasma
luminosity, we sum over individual ionic species. Physically, however,
the line emission is the result of collisions between ions and
electrons (or, in the case of the Bremsstrahlung radiation that
dominates more energetic plasmas, the result of the deflection of
electrons by ions). Whilst the metal ions that dominate the emissivity
of the plasma result from internal, stellar evolution processes, the
left-hand panel of Fig.~\ref{fig:radial_frec_profile} indicates that
the mass density, and therefore the electron density, of the hot CGM is dominated by the metal-poor plasma
accreted from the IGM. We therefore conclude
that the observability (in emission) of the hot CGM of typical
galaxies requires both the outward transport of heavy elements from
the central galaxy, and the accretion of gas from the IGM to compress and heat the recycled gas, and to provide electrons that can excite the metal ions.

\subsection{Iron abundance mass profiles}
\label{sec:Fe_profiles}

\begin{figure}
\includegraphics[width=\columnwidth]{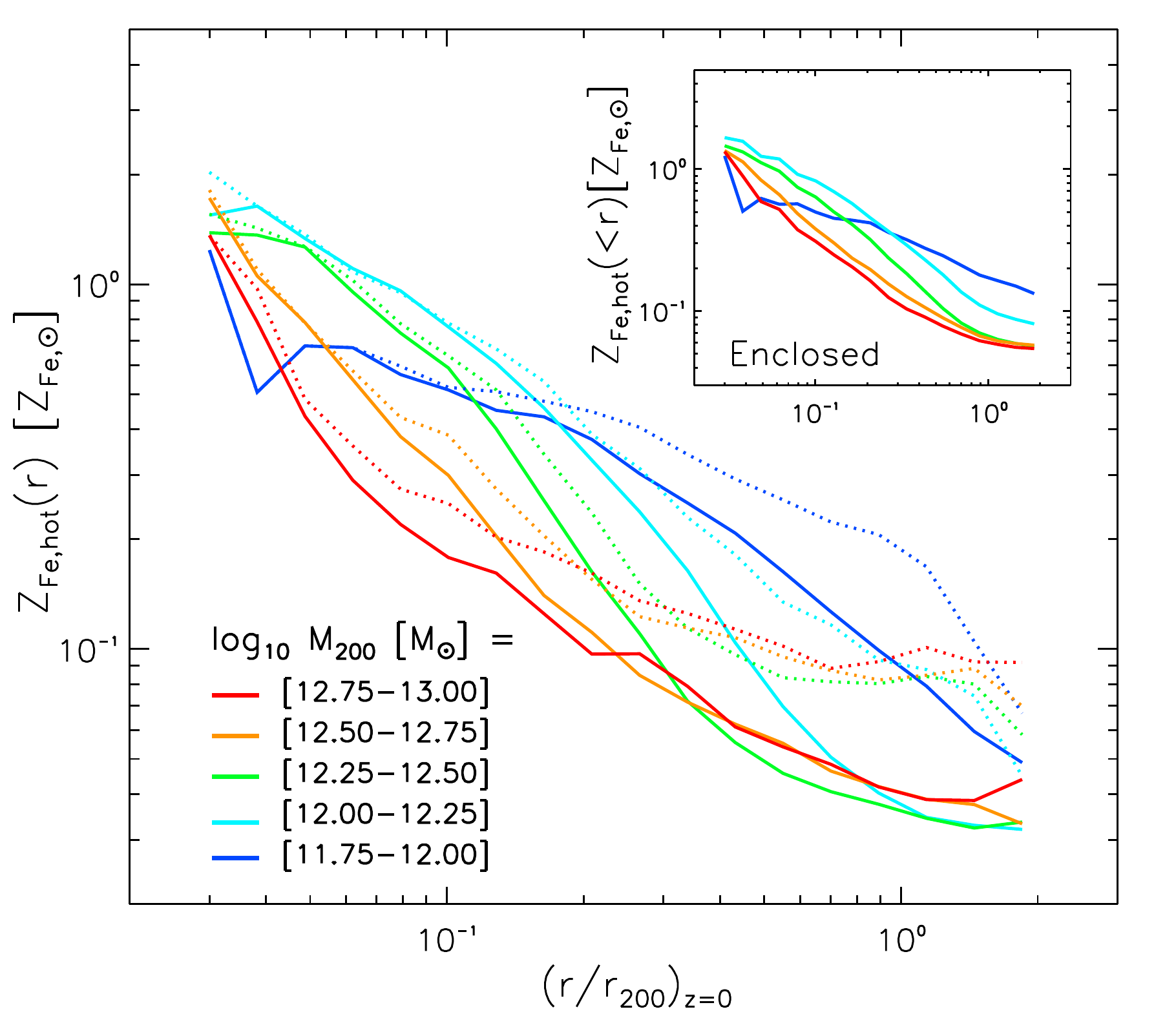}
\caption{Spherically-averaged radial profile of the mass-weighted (\textit{solid lines}) and \lx-weighted (\textit{dotted lines}) iron mass fraction of the hot CGM of \gimic\ galaxies, binned by $M_{200}$ as in  Fig.~\ref{fig:radial_frec_profile}. The profiles of haloes of all masses exhibit a negative gradient. Mildly super-solar abundances are exhibited at the halo centre; the mass-weighted profiles drop to a few percent of solar at the virial radius, whilst the \lx-weighted profiles tend to a tenth of solar, highlighting the presence of a bias in the metallicity measure at large radii. The inset panel shows mean mass-weighted iron mass fraction obtained by integrating over all hot CGM gas within a radius $r$. See text for a discussion of these biases.}
\label{fig:ZFe_profiles}
\end{figure}

The declining contribution of gas recycled by stellar evolution to the
mass of hot CGM at large radii is reflected in the radial profile of its
\textit{mass-weighted} (Eq.~\ref{eq:ZFe_mw}) iron abundance, \ZFe,
plotted with solid lines in Fig.~\ref{fig:ZFe_profiles}. The strong
abundance gradients reflect the centrally-concentrated \ZFe\
distribution visible in the top-right panel of Fig.~\ref{fig:maps}.
The central iron abundance is typically super-solar, but drops to a
few percent of \ZFesun\ at the virial radius ($r_{200}$). Such a
strong metallicity gradient makes estimates of the hot CGM metallicity
dependent on the radius within which it is measured. To illustrate
this, we show in the inset plot the mass-weighted iron abundance
recovered when integrating over the gas enclosed within a given
radius, $Z_{\rm Fe}(<r)$. Within $0.1\,r_{200}$ (approximately $r_{\rm
  e}$ for \lstar\ galaxies), the typical mass-weighted iron
abundance is mildly sub-solar, but drops to $\sim 0.1 \Zsun$ if all
hot CGM gas within $r_{200}$ is considered.

Our discussion so far has focussed on the fact that X-ray measurements of iron abundance are higher than the `true' mass-weighted measurement because the strongly centrally-peaked surface brightness profile of the hot CGM strongly biases the measurement towards the bright, metal-rich centre of the X-ray corona. However, we might reasonably also expect a bias between mass- and \lx-weighted iron abundance measurements \textit{at fixed radius}, because of i) the sensitivity of the plasma emissivity to the presence of metal ions, and ii) the tendency of metals to clump \citep[see e.g.][]{Simionescu_et_al_11_short,Nagai_and_Lau_11,vandeVoort_and_Schaye_13}, either because they induce efficient cooling, and/or because they are associated with dense substructures. Clumping is indeed apparent in the stacked map of hot CGM iron abundance in Fig.~\ref{fig:maps}. To assess the scale of this bias, we plot the \lx-weighted radial profile of \ZFe\ in Fig.~\ref{fig:ZFe_profiles} with dotted lines. At small radii ($r \lesssim 0.2r_{200}$), the mass- and \lx-weighted measurements are similar, but at larger radii the mass-weighted iron abundance exhibits a steeper gradient than the \lx-weighted measurement, such that at $r_{200}$, the latter is a factor of $\sim3$ greater than the former. We conclude then that the effect is significant, but insufficient to explain the overall offset between mass-weighted and \lx-weighted measurements of \ZFe\ when integrated over the entire hot CGM. The combination of a centrally-peaked X-ray surface brightness profile and a strong negative gradient in radial metallicity in the hot CGM is stronger than the bias at fixed radius.

In summary, it is clear that the `true' (i.e. mass-weighted) iron abundance of the hot CGM is significantly lower than the value recovered from an \lx-weighted measurement. The marked difference between the mass-weighted and \lx-weighted iron abundances reconciles the inference of near-solar iron abundances from X-ray spectroscopy of $\sim$\lstar\ galaxies with models of galaxy formation in which the hot CGM is predominantly accreted from the IGM.
The unavoidable \lx-weighting of X-ray data biases the measurements towards the hot, dense and metal-rich gas (with the highest surface brightness) lying close to the galaxy. A mild bias between mass- and \lx-weighted measures exists at fixed radius because of the sensitivity of specific X-ray emissivity to the presence of metal ions. By mass, most hot CGM gas resides at large radius, where there is a greater mass of metal-poor gas, accreted from the IGM, to dilute transported metals. The rapid decline of the X-ray surface brightness of the hot CGM with galactocentric radius seen in Fig.~\ref{fig:maps} highlights why radial metallicity profiles are so challenging to determine from X-ray emission line spectroscopy. We speculate that the best prospect for constraining mass-weighted abundance profiles with existing X-ray instrumentation is to use stacked emission maps or absorption spectroscopy \citep[e.g.][]{Fang_et_al_10}, and to compare these to synthetic observations created with hydrodynamical simulations.

\section{Dissecting the enrichment of galactic coronae}
\label{sec:dissecting_enrichment}

Having established in \S~\ref{sec:coronal_metallicities} that the iron
abundance of the hot CGM associated with $\ga$\lstar\ galaxies in the
\gimic\ simulations is compatible with observational constraints, we
turn to a detailed analysis of how the enrichment proceeds over cosmic
time. We start by examining the location of enriched gas particles at
the epoch of enrichment, and ask whether they were bound to the same
central galaxy they are bound to at $z=0$, a satellite galaxy of the
central galaxy, another galaxy altogether, or bound to no galaxy at
all (i.e. the IGM). We also explore the radial migration of hot,
enriched gas throughout the assembly history of the
galaxies. We subsequently investigate whether the enrichment of the
hot CGM was dominated by prompt Type II SNe detonations or by the slow
release of metals (on $10^9$-year timescales) by Type Ia SNe, AGB
stars, etc.

\subsection{Is the hot CGM enriched internally or externally?}
\label{sec:internal_external}

To assess the relative significance of the reservoirs that contribute
metals to the hot CGM of \lstar\ galaxies at $z=0$, we adopt a similar
methodology to \citet{Font_et_al_11}, who explored the contribution of
various processes to the formation and assembly of stellar haloes. For
each galaxy in our \gimic\ sample, we separate the enrichment history
of its coronal gas into four categories:
\begin{itemize}
\item \textit{in-situ}: an enrichment event is flagged as `in-situ' if, at the time of enrichment, the recipient gas particle was bound to the most massive subhalo of the most massive progenitor (MMP) FoF group of the galaxy at $z=0$.
\item \textit{satellite of MMP}: the event is flagged as `satellite enrichment' if, at the time of enrichment, the recipient gas particles was bound to any other subhalo of the MMP.
\item \textit{external galaxy}: the event is flagged as `external
  enrichment' if, at the time of enrichment, the recipient gas
  particle was bound to any subhalo of any FoF group other than the
  MMP. 
\item \textit{smooth}: the event is flagged as `smooth' if, at the time of enrichment, the recipient gas particle was not part of any FoF group. 
\end{itemize}
\noindent The categories are unambiguous insofar that they account for
all enrichment events and are mutually exclusive. 

The procedure by which we identify the MMP and classify enrichment
events is as follows. For a given central galaxy at $z=0$, we select
all dark matter particles within $r_{200}$ belonging to the dominant
(most massive) subhalo. We use the unique IDs assigned to these
particles to identify the FoF group in previous snapshots that
contains the greatest number of these particles. This FoF group is
said to be the MMP and the most massive subhalo of that FoF group is
assumed to be the MMP of the dominant subhalo of the system at $z=0$.
Each of the gas particles that make up the hot CGM at $z=0$ is tracked
back in time through all previous simulation outputs (`snapshots'),
and the iron mass added to the particle is assigned to one of the four
categories above, based on which, if any, FoF group and subhalo the
gas particle belonged to at that time.

Each \gimic\ volume has approximately 60 snapshots, spanning the
redshift range $z=20$ to $z=0$. The time of each enrichment event will
not, in general, correspond to the redshift for which the snapshot
data were written. We therefore must use the snapshot that is closest
in time to identify where the gas particle was enriched. This could,
in certain circumstances, lead to a misidentification of the location
of the enrichment (i.e. to which, if any, subhalo it belonged when it
was enriched). However, we have explicitly checked that this is not an
important issue by re-running our tracing algorithm using only every
second snapshot, and find that the in-situ enrichment fraction
recovered in this case does not differ significantly; we therefore
conclude that the redshift sampling of our snapshot data is sufficient
to deduce whether a given particle was enriched `internally' (i.e.
in-situ) or `externally' (i.e. in a satellite, an external galaxy, or
the IGM). We trace each of the $z=0$ coronal gas particles back in
time as far as $z=10$ to determine their enrichment category. We
ignore the very small fraction of metal mass synthesised at $z>10$.

\begin{figure}
\includegraphics[width=\columnwidth]{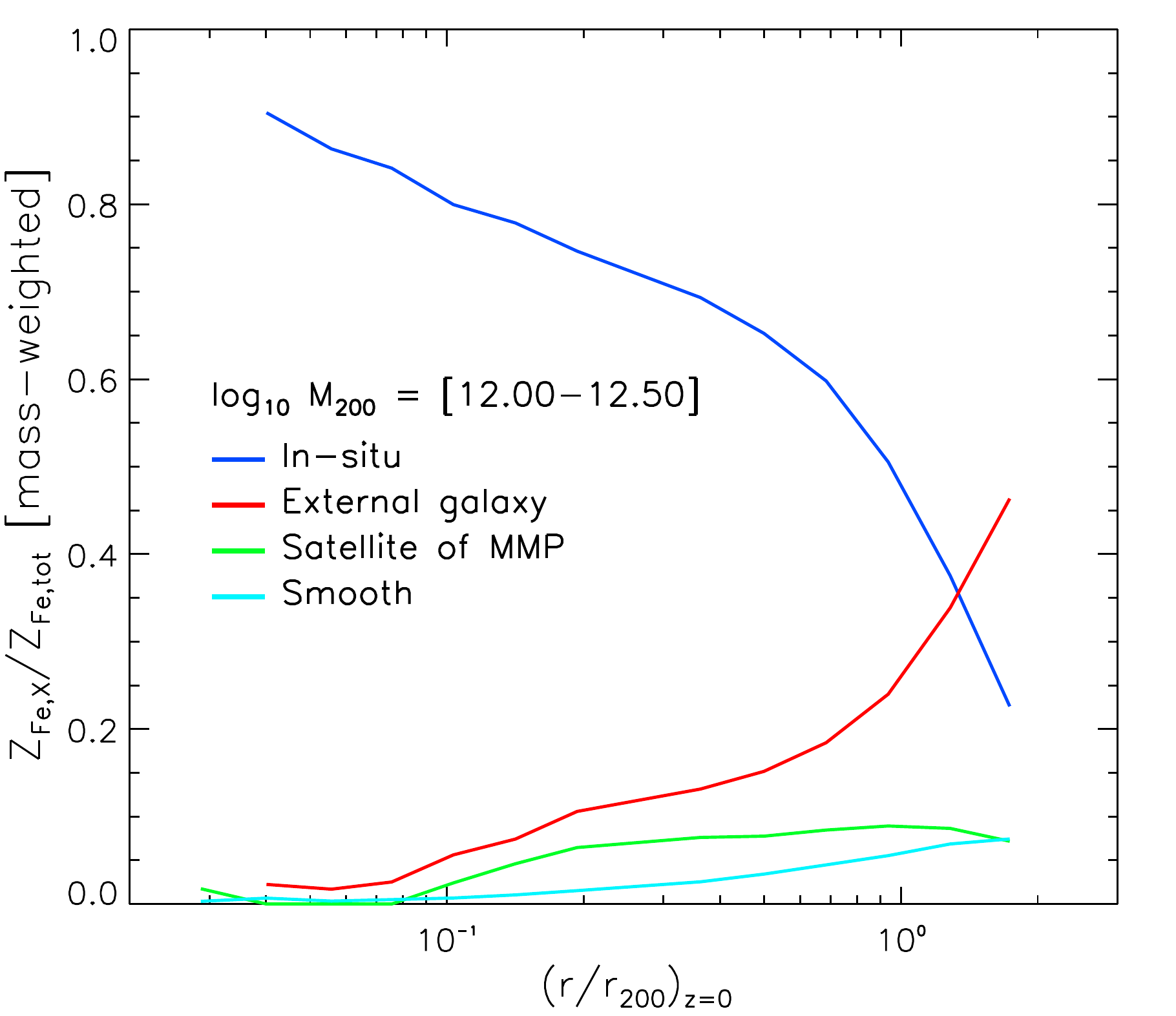}
\caption{Spherically-averaged mass-weighted radial profiles of the coronal iron mass fraction contributed by the categories of enrichment described in \S~\ref{sec:internal_external}: in-situ (\textit{blue}), within an external galaxy (\textit{red}), within a satellite of the central galaxy (\textit{green}), and external to any FoF halo (`smooth', \textit{cyan}). Within $r_{200}$, in-situ enrichment dominates, indicating that the majority of coronal metals are transported from the stars comprising the central galaxy. At large radii, where relatively few in-situ metals reside, metals aqcquired from external galaxies (e.g. by winds or ram-pressure stripping) become significant, but not dominant. Satellites of the main progenitor of the galaxy contribute only a small fraction of the total metal content of the hot CGM.}
\label{fig:mode_ratios}
\end{figure}

In Fig.~\ref{fig:mode_ratios}, we plot the fractional contribution of
these categories to the overall iron mass fraction radial profile, for
the $\sim350$ galaxies with halo mass in the range $10^{12.00} <
M_{200}/{\rm M}_\odot < 10^{12.50}$. Within the virial radius in-situ
enrichment dominates, indicating that the majority of coronal metals
are synthesised by, and transported from, stars comprising the central
galaxy. At large radii ($\sim r_{200}$), the `external galaxy'
category becomes significant (but not dominant), indicating that
coronal outskirts are, in part, enriched by the transfer of gas from
external galaxies to the MMP, for example via ram-pressure stripping
or entrainment within supernova-driven (SN-driven) outflows. It is clear that the contribution of external galaxies to the metal mass of hot gas confined by the potential of haloes will be a function of halo mass, since the central galaxy of galaxy groups and clusters hosts a decreasing fraction of the overall stellar mass in more massive haloes \citep{Lin_and_Mohr_04}. However, even in haloes as massive as $\sim 10^{14}\Msun$ the central galaxy typically comprises over 50 percent of galactic stellar mass bound to the halo, so the curves presented in Fig.~\ref{fig:mode_ratios} are unlikely to be particularly sensitive to $M_{200}$.

The enrichment of gas associated with satellites of the MMP
contributes, on average, only a small fraction of the total iron mass
of the simulated coronae. This is, in part, due to our sample
selection criterion that the main subhalo of a FoF group should
account for at least 90 percent of the virial mass, a criterion we
imposed to ensure that we study the hot coronal gas associated with
galaxies rather than major mergers or galaxy groups. However, we have
confirmed that this bias is small by relaxing this criterion to
require that $M_{\rm sub} > 0.5 M_{\rm FoF}$, which produces very
similar results.

The contribution to the overall fraction of coronal iron from intergalactic gas (`smooth') can be viewed as an uncertainty estimate on the plotted fractions. Since \gimic\ assigns the metals synthesised by star particles to their neighbouring gas particles (i.e. those within an SPH smoothing kernel), in the limit of infinite resolution nearly all gas particles should be enriched as part of a FoF ($b=0.2$) group (the fraction of truly intergalactic stars is very small). However, the finite particle resolution of the simulation means that it is possible for the smoothing kernel of star particles to enclose a small fraction of unbound particles. Fortunately, this issue is most problematic in ill-resolved, low-mass galaxies, which synthesise only a small fraction of the cosmic heavy elements, as confirmed by Fig.~\ref{fig:mode_ratios}. We therefore neglect this component in subsequent plots. We note that our findings are qualitatively consistent with those of \citet{Shen_et_al_12}, who simulated the formation of a massive galaxy to $z=3$ and found that the central galaxy was the dominant source of circumgalactic metals, with only $\sim10$ percent contributed by satellite companions at that epoch.

In addition to ascertaining the mode by which gas is enriched, we can obtain a simple quantitative diagnostic that elucidates \textit{where} coronal gas was enriched by defining a quantity for each gas particle that we term the `iron enrichment weighted radius', $r_{\rm Fe}$:
\begin{equation}
r_{\rm Fe} = \frac{\sum_i{\Delta m_{i,{\rm Fe}}r_{i}}}{\sum_i{\Delta m_{i,{\rm Fe}}}},
\label{eq:iron_weighted_radius}
\end{equation}
\noindent where $i$ sums over all enrichment events, $\Delta m_{i,{\rm Fe}}$ is the growth in iron mass of the gas particle per event, and $r_{i}$ is the radial distance (in physical coordinates) of the enriched gas particle from the halo centre at the time of enrichment \citep[see also][]{Wiersma_et_al_10,Shen_et_al_12}. This quantity is not followed explicitly on a timestep-by-timestep basis in \gimic, hence we compute it in post-processing from snapshots.

\begin{figure}
\includegraphics[width=\columnwidth]{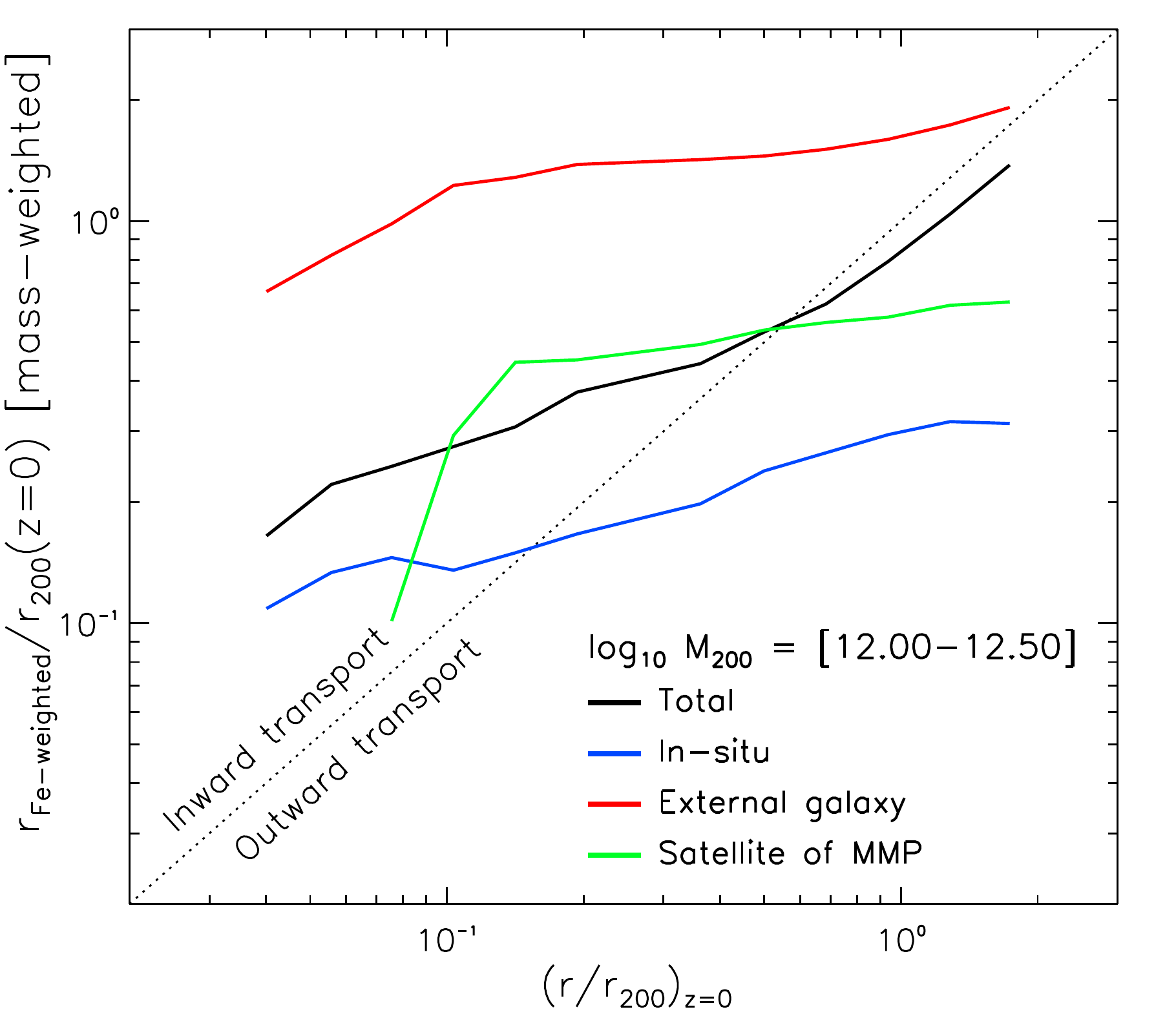}
\caption{Spherically-averaged mass-weighted radial profile of the iron enrichment weighted radius (Eq.~\ref{eq:iron_weighted_radius}) of coronal gas at $z=0$, split by the enrichment categories described in \S~\ref{sec:internal_external}. The diagonal dotted line marks the locus $r_{\rm Fe} = r(z=0)$, denoting no radial migration. The mode of iron synthesis correlates strongly with the galactocentric radius at which it was synthesised, and both inwards and outwards metal transport are significant. Overall, outward transport dominates, since most coronal iron is produced in-situ and re-distributed by galaxy-scale winds and convection.} 
\label{fig:migration}
\end{figure}

In Fig.~\ref{fig:migration}, we plot $r_{\rm Fe}$ as a function of $z=0$ radius, binned by the enrichment categories defined above. Iron synthesised within external galaxies (\textit{red}) was, by definition, synthesised beyond the virial radius of the MMP, but later became bound to the potential of the central galaxy-halo system. This can occur as a result of metal transport due to winds driven by the external galaxy \citep[e.g.][]{Shen_et_al_12}, or, if the external galaxy subsequently merges with the MMP, due to ram pressure stripping of the external galaxy's CGM as it interacts with the hot CGM of the MMP \citep[][]{Abadi_Moore_and_Bower_99,Bahe_et_al_12}. Externally synthesised iron exhibits a mild correlation between the $r_{\rm Fe}$ and $r(z=0)$, such that the iron synthesised closer to the coronal centre resides closer to the galactic centre at $z=0$. 

The dominant in-situ (\textit{blue}) mode of iron synthesis exhibits a significant correlation between $r_{\rm Fe}$ and $r(z=0)$. Interestingly, this curve crosses the dotted line in Fig.~\ref{fig:migration} that marks the locus $r_{\rm Fe} = r(z=0)$, and hence the in-situ iron is characterised by both inward and outward transport. The inward transport is confined to iron synthesised within $\sim0.1$ of the present day virial radius of galaxies, which roughly corresponds to the scale length of star forming galaxy discs: this iron was likely therefore synthesised in the outskirts of galaxy discs and subsequently migrated radially in the disc plane. However, most in-situ iron is transported away from the central galaxy during its assembly history. We therefore conclude that whilst both internal and external proceses contribute to the enrichment of the hot CGM associated with \lstar\ galaxies at $z=0$, it is the internal, in-situ category that dominates.

\subsection{When were coronae enriched?}
\label{sec:when_enriched}

\begin{figure}
\includegraphics[width=\columnwidth]{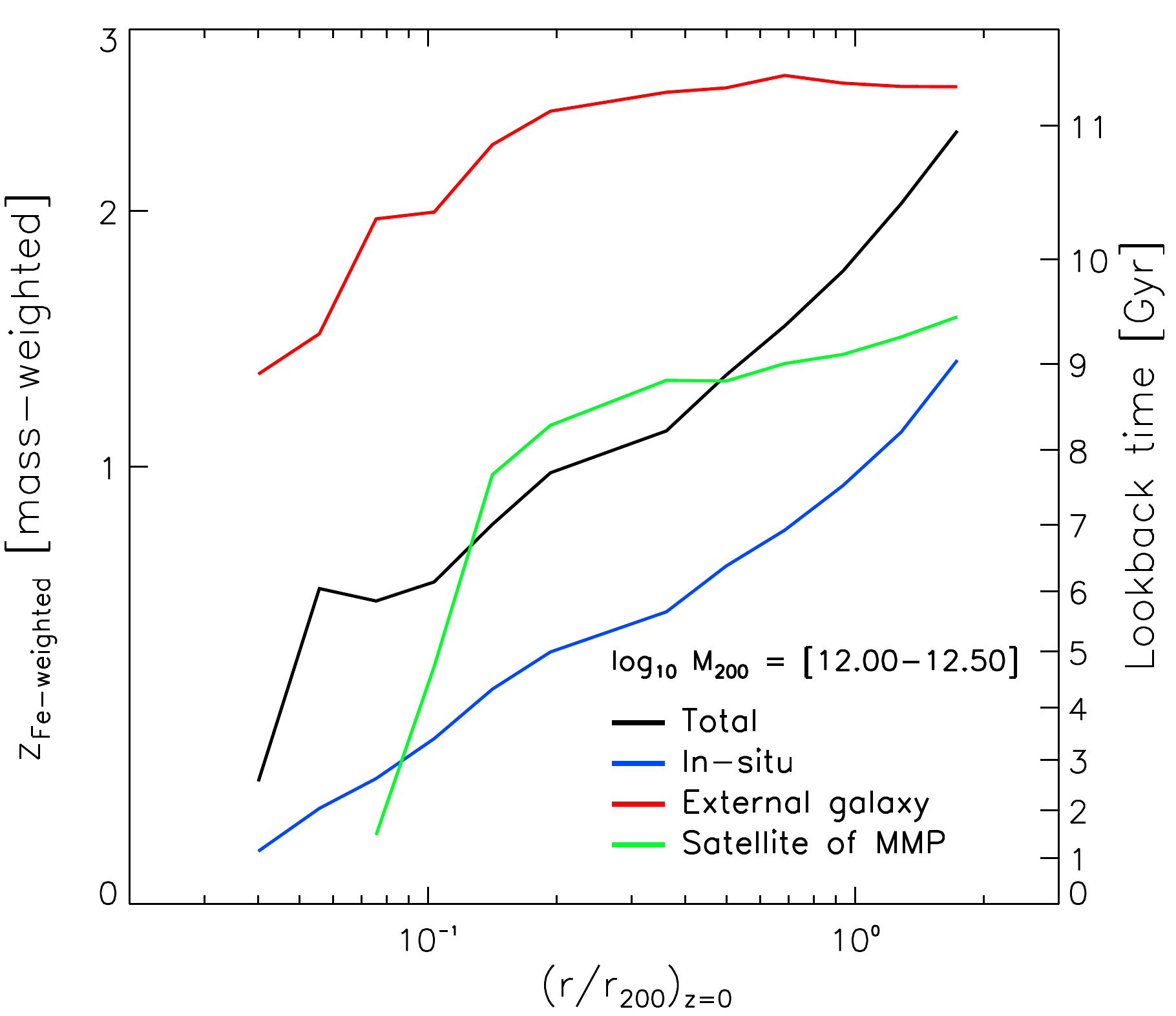}
\caption{Spherically-averaged mass-weighted radial profile of the iron
  enrichment weighted redshift (Eq.~\ref{eq:iron_weighted_redshift})
  of coronal gas at $z=0$, split by the enrichment categories
  described in \S~\ref{sec:internal_external}. The enrichment
  categories differ markedly, with enrichment by external galaxies
  occuring at relatively early times and in-situ enrichment being
  dominant at late epochs. All categories exhibit a trend such that
  the metals located farthest from the galactic centre were enriched
  at the earlist epochs. We interpret this as a signature of a travel-time, such that the metals synthesised earliest have been in transit from the galactic centre for the longest time.}
\label{fig:iron_weighted_redshift}
\end{figure}

In analogy with the iron mass weighted radius defined in Eqn.~\ref{eq:iron_weighted_radius}, we can we can obtain a simple quantitative diagnostic that elucidates \textit{when} coronal gas was enriched by defining an `iron mass weighted redshift', $z_{\rm Fe}$, for each particle \citep[see also][]{Wiersma_et_al_10}:
\begin{equation}
z_{\rm Fe} = \frac{\sum_i{\Delta m_{i,{\rm Fe}}z_{i}}}{\sum_i{\Delta m_{i,{\rm Fe}}}},
\label{eq:iron_weighted_redshift}
\end{equation}
\noindent where, $i$ once again represents an enrichment event, $\Delta m_{i,{\rm Fe}}$ is the growth in iron mass of the gas particle per event, and $z_{i}$ is the epoch of enrichment. As for $r_{\rm Fe}$, we compute this quantity in post processing and, in analogy with Fig.~\ref{fig:migration}, plot $z_{\rm Fe}$ as a function of $z=0$ radius, binned by the enrichment categories defined in \S~\ref{sec:internal_external}, in Fig.~\ref{fig:iron_weighted_redshift}. The enrichment categories also exhibit significant differences in this diagnostic. 

Iron contributed by external galaxies was synthesised at early times
(typically $1\lesssim z \lesssim 3$), and a mild correlation between
$r(z=0)$ and $z_{\rm Fe}$ is apparent, such that the
externally-synthesised iron close to the coronal centre at $z=0$ was
synthesised more recently than that in the coronal outskirts. The
in-situ iron exhibits a strong correlation between $r(z=0)$ and
$z_{\rm Fe}$, such that the most-recently synthesised iron is closest
to halo centre. In-situ iron found within $0.1r_{200}$ was typically
synthesised within the last $3-4\Gyr$, whilst the in-situ iron at
$\sim r_{200}$ was synthesised $7-8\Gyr$ ago. This trend is a clear
signature of the operation of metal transport, insofar that the iron
synthesised at the earliest epochs has had the longest period of time
to be transported to large radii. Interestingly, qualitatively similar
trends operate for the satellite and external galaxy categories,
indicating that the majority of metals synthesised in satellite and
external galaxies are likely deposited close to the halo centre, and
transported outwards in the same fashion as the in-situ iron. This
result is in qualitative agreement with \citet{Wiersma_et_al_10}, who
found that metals residing in low density gas were typically ejected
at early times, by low mass haloes.

\subsection{How are the metals transported?}
\label{sec:metal_transport}

Having established that the majority of metals in the hot CGM of today's \lstar\ galaxies were synthesised at the halo centre and transported outwards during the assembly history of the galaxy, it is interesting to consider briefly how the transport proceeds. The entrainment of metals in galaxy winds is a natural mechanism to consider, since the winds originate at the galactic centre and propagate into the CGM. Moreover, metals entrained in outflows are readily observed at both high and low redshift \citep[e.g.][]{Heckman_Armus_and_Miley_90,Pettini_et_al_01,Veilleux_Cecil_and_Bland-Hawthorn_05}. The importance of metal transport by winds as a means to enrich circum- and intergalactic gas was recently studied in detail by \citet{Wiersma_Schaye_and_Theuns_11}, who used cosmological hydrodynamical simulations to show that winds are the primary mechanism by which the cold-warm IGM is enriched.

However, we demonstrated in Paper I that the hot CGM of \lstar\
galaxies in the \gimic\ simulations is dominated at late-times by
quasi-hydrostatic gas that has never been part of the ISM (and hence
has never been ejected in a galactic fountain). It is therefore unlikely that outflows are the \textit{dominant} metal transport mechanism at late times ($z\lesssim1$) in \gimic. There is, however, another mechanism
by which the metals can be transported: convection.
\citet{McCarthy_et_al_07b_short} demonstrate that hydrostatic haloes
grow in an inside-out fashion, such that the orbital kinetic energy
associated with infalling gas is thermalised within the hot CGM of the
main progenitor. The entropy of the infalling gas itself is largely
unaffected, so it sinks to the halo centre and thus steadily convects
the higher entropy and enriched gas, initially at the halo centre, to
larger radii. It follows naturally that the metals synthesised
earliest will reside furthest from the halo centre. Although some
fraction of the hot CGM will flow inwards due to radiative cooling, a
significant fraction of the gas has a long cooling time and will
therefore convect as the halo is assembled. The dominance of this
mechanism on timescale grounds appears plausible; transporting metals
a distance of 250\kpc\ (the typical $r_{200}$ at $z=0$ for a galaxy
in our simulation sample) in a period of $8\Gyr$ requires a net
outflow velocity of only $\sim30\kms$, roughly an order of magnitiude
lower than the characteristic velocity of SN-driven outflows as they emerge from the ISM.

\subsection{Which mechanism dominates iron synthesis?}
\label{sec:which_mechanism}

Having analysed the radial distribution of iron in the hot CGM, and
the nature of its release into, and transport within, the hot CGM, we now turn to the
\textit{source} of the iron. We consider the synthesis by Type II SNe
that occurs promptly ($\lesssim30$\,Myr) after star formation
episodes, and the synthesis by Type Ia SNe that is delayed with
respect to star formation. As described by \citet{Wiersma_et_al_09},
the chemodynamics implementation adopted by \gimic\ also accounts for
the fact that, whilst AGB stars do not synthesise iron, they do
`lock-up' significant quantities of iron when an SSP forms from
pre-enriched gas. This iron is then gradually released by AGB envelope
shedding on $\sim$\Gyr\ timescales.

\begin{figure}
\includegraphics[width=\columnwidth]{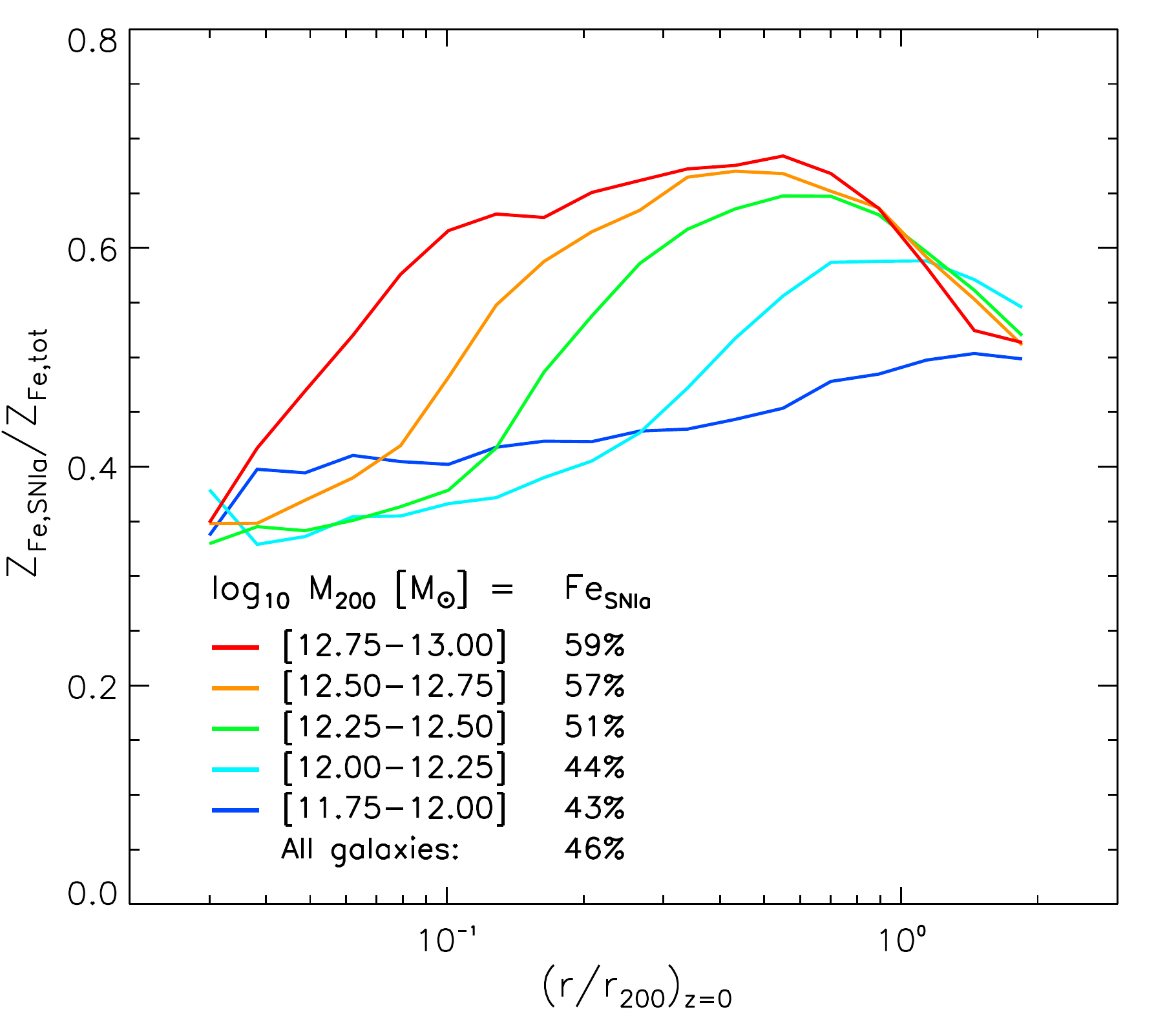}
\caption{Spherically-averaged radial profile of the mass fraction of iron within the hot CGM, synthesised by Type Ia SNe. The median iron mass fraction synthesised by Type Ia SNe within \rhalo\ for each halo mass bin is quoted next to the legend. Iron associated with the hot CGM of \gimic\ galaxies is predominantly synthesised by Type Ia SNe; this is in spite of our adopted yields and Type Ia SN rates being such that an evolved ($t_{\rm age} > 10^9\Gyr$) SSP with a Chabrier IMF tends to a near-equipartition of iron production from Type II SNe and Type Ia SNe. We infer that iron synthesised by Type II SNe is preferentially i) entrained in outflows and ejected from the system, or ii) locked up in secondary generations of stars.}
\label{fig:ZFe_SNe_profiles}
\end{figure}

In Fig.~\ref{fig:ZFe_SNe_profiles}, we split the profiles presented in
Fig.~\ref{fig:ZFe_profiles} into contributions from the two types of
SNe, and plot the spherically-averaged radial profile of the Type
Ia-to-total iron mass fraction. The radially-integrated
($0<r<$\,\rhalo) iron mass fraction synthesised by Type Ia SNe is
quoted for each halo mass bin in the legend. Both types of SN
contribute at all radii for galaxies of all masses. The approximate
equipartition of iron synthesised by each SN type follows from the
behaviour of the stellar populations comprising the galaxy associated
with each corona; adopting the yields and Type Ia SNe rates presented
by \cite{Wiersma_et_al_09}, an evolved ($t_{\rm age}\gtrsim 10^9\yr$)
SSP with a Chabrier IMF tends to a broad equipartition of the iron
mass fraction produced by each type of SN, particularly if the SSP
has sub-solar metallicity. However, we caution that this statement is
sensitive to the assumed cosmic Type Ia rate\footnote{The SN Ia delay
  function is ill-constrained by theory and observation. The \gimic\
  simulations adopt an $e$-folding delay function with a
  characteristic timescale of $\tau=3.3\Gyr$.} and the iron yield of Type Ia and II SNe.

There is, however, a mild but significant preference for iron
synthesised by Type Ia SNe over that from Type II SNe in all but the
least massive galaxies. We deduce two reasons why iron synthesised by
Type II SNe is inhibited from residing in the hot CGM of galaxies at
late cosmic epochs: firstly, the iron sythesised by Type II SNe is
preferentially entrained in the energetic outflows that these events
trigger, enabling a significant fraction of iron synthesised in this
fashion to be ejected into the IGM at early epochs\footnote{This is
  clearly a significant process, since the mass of gas ejected by a
  galaxy's stars over cosmic history is typically comparable to, or
  greater than, the mass of gas associated with the galaxy's potential
  at $z=0$.}. Secondly, because Type II SNe synthesise iron promptly,
any iron from Type II SNe that is not ejected from the galaxy's
potential is deposited into the dense ISM, where the enrichment products trigger efficient
radiative cooling and readily become locked-up in secondary
generations of stars. This is reflected in the abundance patterns of
stars in the simulated galaxies: the fraction of Type II
SN-synthesised iron in the stellar component of galaxies comprising
our sample is typically 70-80 percent \citep[see
also][]{Font_et_al_11}. However we caution that this is likely a mild
overestimate of the true fraction, since the inefficiency of feedback
in the most massive galaxies comprising our sample will lead to too
high a fraction of Type II SN-synthesised iron being unable to escape
the confines of the ISM.

The trend with halo mass, such that the hot CGM of the most massive
haloes is more strongly dominated by iron synthesised by Type Ia SNe
($Z_{\rm Fe,SNIa}/Z_{\rm Fe,tot} \sim 60$ percent) can be interpreted
in terms of the different star formation histories of the galaxies
associated with these coronae. As shown in C09 (see Fig. 6 therein),
more massive galaxies in \gimic\ complete a greater fraction of their
star formation at high-redshift, enabling a greater fraction of their
stellar populations to evolve to the stage where a significant
fraction of their ejected iron is contributed by Type Ia SNe.

The coronal iron mass fraction from Type Ia SNe in \gimic\ is slightly
lower than the $70-90$ percent estimated by HB06 for their sample of
ellipticals, whose masses and X-ray luminosities are consistent with
the most massive galaxies in the \gimic\ sample. However, comparison
of this result is not entirely instructive; we have not limited the
\gimic\ sample to `elliptical' galaxies, and the limited X-ray
spectroscopy available for disc galaxies with high star formation
rates unsurprisingly indicate abundance patterns more consistent with
Type II SN yields \citep{Richings_et_al_10}. \citet{Tang_and_Wang_10}
argue that the observable iron content of the hot CGM might in fact
provide an unpresentative view of the total mass of iron synthesised
by Type Ia SNe, because the iron is likely to be entrained in very
hot, tenuous ejecta with a low specific soft X-ray emissivity and a large
buoyancy that results in its rapid transport to large radii. Perhaps
more importantly, however, SN nucleosynthesis yields remain uncertain at the factor $\sim2$ level, and the cosmic Type Ia SN rate remains poorly constrained by both theory and observation. The simulations could remain entirely compatible with the cosmic Type Ia SN rate using double the adopted rate, and this would unquestionably improve the correspondence between the observed and predicted Type Ia iron fraction in the hot CGM. For this reason, we caution that comparison of the simulations with the currently available observational constraints is not particularly instructive.

\section{Summary and discussion}
\label{sec:discussion}

We have investigated the enrichment of the hot circumgalactic medium
(CGM) associated with $\sim$\lstar\ ($10^{10} \lesssim M_\star \
\lesssim 10^{11.5} \Msun$) galaxies at $z=0$, using the
\textsc{Galaxies-Intergalactic Medium Interaction Calculation} 
\citep[\gimic;][]{Crain_et_al_09_short}. These simulations produce a
large sample ($n = 617$) of well-resolved galaxies at $z=0$, with a
distribution of morphologies that broadly corresponds to that observed
in the local Universe. We showed in \citet[][Paper I]{Crain_et_al_10} that the
simulations reproduce the observed $z=0$ scalings of X-ray luminosity
with $K$-band luminosity, star formation rate, and disc rotation
velocity. The simulations also reproduce the observed $z=0$ stellar
mass - rotation speed (or `Tully-Fisher') and stellar mass - halo mass
relations for $10^9 \lesssim M_\star/\Msun < 10^{10.5}$, but we
caution that they still suffer from some overcooling for $M_\star \ga
10^{11} \Msun$ \citep{McCarthy_et_al_12b}. The results from our simulations are summarised as follows:

\begin{itemize}

\item The hot CGM is dominated in a mass-weighted sense by metal-poor gas accreted from the intergalactic medium. Therefore the iron abundance of the hot CGM, when integrating over all gas within the virial radius, is typically $\lesssim 0.1{\rm Z}_{\rm Fe,\odot}$.

\item Conversely, the X-ray emissivity of the hot CGM is dominated by line emission from collisionally excited metal ions that were synthesised in stars and transported out of the ISM. However, the excitation is caused by collisions with electrons; since heavy elements make only a small contribution to the free electron budget, these electrons are primarily sourced from the metal-poor gas accreted from the IGM.

\item The iron found in the hot CGM of galaxies in the \gimic\
  simulations at $z=0$ was predominantly injected into gas that was,
  at the epoch of enrichment, already bound to the most massive
  progenitor of the galaxy, a process we term `in-situ' enrichment. 

\item We find a clear correlation between the galactocentric radius of metals at $z=0$, and the redshift at which they were synthesised, such that those synthesised at earlier cosmic epochs
  reside farther from the galaxy. Enrichment therefore proceeds
  in an `inside-out' fashion, requiring the transport of metals from
  the central galaxy into the CGM. We speculate that this transport is
  dominated by SNe-driven winds at early times, and by convection
  associated with the inside-out growth of the halo at later times,
  when the hot CGM is quasi-hydrostatic.

\item Inside-out enrichment establishes a strong negative iron abundance gradient in the hot CGM. The spherically-averaged radial profile of the iron abundance is typically super-solar at small radii ($\lesssim 0.05r_{200}$), and rapidly declines to only a few percent of the solar iron abundance at $r_{200}$, or to about ten percent of solar if the metallicity is weighted by the luminosity. .

\item The strong negative metallicity gradient established by inside-out enrichment concentrates the majority of the metal ions associated with the hot CGM inside a relatively small radius. In contrast, following simple volumetric arguments, the majority of the mass of the hot CGM resides at large radius. Therefore, X-rays do not trace the distribution of the hot CGM in a simple fashion, rendering metallicity measurements of the hot CGM inferred from spatially unresolved X-ray spectroscopy biased towards the metal-rich gas close to the galaxy. 

\item In addition, the sensitivity of plasma emissivity to the presence of metal ions induces a bias such that the luminosity-weighted metallicity of the hot CGM at a fixed radius is typically elevated with respect to its mass-weighted metallicity at the same radius. This induces a bias between luminosity- and mass-weighted metallicity measurements even when inferred from spatially-resolved X-ray spectroscopy.

\item Both biases lead to abundance measures that are elevated with respect to the `true', mass-weighted, measure. Therefore, despite exhibiting a typical iron abundance of $\lesssim
  0.1{\rm Z}_{\rm Fe,\odot}$ at $z=0$, the median \lx-weighted iron
  abundance of the hot CGM is approximately ${\rm Z}_{\rm Fe,\odot}$,
  with a $2\sigma$ scatter of approximately one decade, across 5
  decades in \lx.
\end{itemize}

The simulations yield a correlation between \ZFe\ and
\lx\ that is weaker than implied by the combination of the
\citet{Humphrey_and_Buote_06} and \citet{Athey_07} datasets, but the recovered iron abundances are broadly consistent
(within the 2$\sigma$ scatter) with these data. The precise nature
and origin of this relation is clearly an avenue of interesting future
theoretical and observational enquiry. Similarly, the median core radius of our hot CGM profiles is
larger than the core sizes inferred from the small number of measurements
inferred from deep exposures of individual galaxies, or stacked profiles extracted from
the \rosat\ All-Sky Survey. As discussed in \S~\ref{sec:stacks}, we do
not consider this a crucial shortcoming of our model, but note that the establishment of the surface density profile of
circumgalactic gas remains an unresolved astrophysical problem that is
well suited for study with the next generation of hydrodynamical simulations.

\subsection{The interpretation of X-ray data in the context of galaxy formation theory}

As remarked in \S~\ref{sec:observational_data}, a central aim of this
study is to ascertain whether the general picture of the formation of
the hot CGM that is predicted by galaxy formation theory, and seen in
cosmological hydrodynamical simulations, is compatible with constraints
inferred from the most authoritative X-ray datasets presently
available. 

The \gimic\ simulations follow the hierarchical assembly of the galaxies and include detailed phenomenological models for the formation of stars, the effects of SN-driven feedback, and the chemical evolution of evolving stellar populations.  Whilst the simulations are known to produce inaccurate results in certain regimes \citep[e.g. C09,][]{McCarthy_et_al_12b} and necessarily offer an incomplete description of the formation of galaxies, they can be considered an authoritative testbed for confrontation with observational measurements, by virtue of the broad range of observable properties of local $\sim$\lstar\ galaxies that they successfully reproduce \citep{Font_et_al_11,McCarthy_et_al_12a,McCarthy_et_al_12b}.

In Paper I and in this study, we establish that $\sim$\lstar\ galaxies in \gimic\ broadly reproduce the key observed X-ray scaling relations of galaxies (i.e. with rotation velocity, stellar mass and star formation rate), and their inferred luminosity-weighted hot CGM iron abundance. Although the simulations are formally not fully consistent with the HB06 and Ath07 datasets, there is a broad correspondence between the two. This is a significant result, since the relatively high metallicity of the hot CGM of local galaxies has been perceived as a challenge to the prevailing model of galaxy formation, within which the hot CGM is assumed to form primarily via the accretion of metal-poor gas from the intergalactic medium throughout the assembly of the galaxy. 

Our simulations indicate that the hot CGM of \lstar\ galaxies is indeed dominated by gas accreted from the IGM. This picture appears similar to that advanced by simple analytic/semi-analytic models \citep[e.g.][]{White_and_Rees_78,White_and_Frenk_91}, but in fact differs in two important ways. Firstly, in contrast to the common assumption that the radial density profile of the hot CGM traces that of the underlying dark matter distribution, we showed in Paper I that SN-driven feedback lowers the normalisation of the density profile and acts to soften the central cusp into a core by preferentially ejecting low entropy gas from the halo centre. The reduction in central gas density lowers the luminosity of the halo by up to two orders of magnitude, because both line emission from collisional excitation and Bremsstrahlung radiation vary with the square of the gas density.

Secondly, as shown here, the dynamically complex and time-dependent enrichment of the hot CGM by the central galaxy, its satellites and external galaxies, establishes a strong negative radial metallicity gradient. This contrasts with the assumption of complete, uniform mixing within the cooling radius adopted by WF91. Since the emissivity of the gas is particularly sensitive to the presence of metal ions, this is an important difference that, like the density profile, has a marked effect on the luminosity and surface brightness profile of the hot CGM. Therefore, it is particularly encouraging that, in contrast to WF91, \gimic\ broadly reproduces the luminosity \textit{and} surface brightness profiles of galaxies in the local Universe. 

This notwithstanding, the interpretation of diffuse X-ray emission
associated with local $\sim$\lstar\ galaxies remains controversial:
the most common interpretation in the literature is that the
emission associated with disc galaxies is a signature of hot plasma
entrained in outflows driven by Type II SNe
\citep[e.g.][]{Strickland_et_al_04,Wang_05,Tullmann_et_al_06,Rasmussen_et_al_09},
whilst that associated with ellipticals is due to the heating of cold
ISM gas by its interaction with the ejecta of intermediate age (AGB)
stars and Type Ia SNe
\citep[e.g.][]{Mathews_90,Ciotti_et_al_91,Mathews_and_Brighenti_03,Parriott_and_Bregman_08}. Whilst
simple models based on these premises can lead to suggestive links
between the optical and X-ray properties of galaxies, such models are,
by construction, incomplete, because they start from idealised initial
conditions, and are often unphysical, for example when assuming that galaxies are hydrodynamically decoupled from the intergalactic medium. Moreover, such models neglect, by construction, the role of the CGM as a reservoir of fuel for on-going star formation in galaxies. As we have argued in this study, a more complete model of galaxy formation is necessary to study the X-ray emission of the hot CGM, because it appears that the emission is the result of both accretion from the IGM and the transport of heavy elements from the ISM into the hot CGM. 

The outstanding challenge is therefore to accomodate the
interpretation of X-ray observations of galaxies within the context of
modern hierarchical galaxy formation theory. As demonstrated in Paper
I and in this study, many such observations are in fact reproduced by
realistic, self-consistent hydrodynamical simulations that adopt
cosmological initial conditions. For example, the broad correlations
between \lx\ and \sfr\ (in disc galaxies) and \mstar\ (in ellipticals)
that provide the fundamental motivation for the idealised models
described above, arise naturally in the \gimic\ simulations because
\lx, \sfr\ and \mstar\ all correlate with \mhalo\ for $\sim$\lstar\
galaxies. It seems not unreasonable to expect, therefore, that improved
modelling in this context will enable future simulations to reproduce
a wide range of X-ray properties and scaling relations accurately. Moreover, as remarked in Paper I, the fundamental nature of the hot CGM of typical galaxies is, in principle, directly accessible via X-ray line diagnostics. Future X-ray observatories with the ability to obtain high-sensitivity spectroscopy at high spatial and spectral resolution therefore offer a direct means to identify the fraction of hot CGM entrained in outflows or residing in quasi-hydrostatic reservoirs. Observations with these facilities may thus verify or falsify the key role in the formation and evolution of galaxies that modern theoretical galaxy formation models assign to hot gaseous coronae.


\section*{Acknowledgements}  
\label{sec:acknowledgements}

The authors thank the anonymous referee for insightful
suggestions that improved the manuscript. IGM is supported by an STFC Advanced Fellowship. TT acknowledges the
hospitality of the Kavli Institute for Theoretical Physics, Santa
Barbara. CSF acknowledges a Royal Society Wolfson Research Merit
Award. The simulations presented here were carried out by the Virgo
Consortium for Cosmological Supercomputer Simulations using the HPCx
facility at the Edinburgh Parallel Computing Centre (EPCC) as part of
the EC’s DEISA `Extreme Computing Initiative', the Cosmology Machine
at the Institute for Computational Cosmology of Durham University, and
on the Darwin facility at the University of Cambridge. This research
was supported in part by the National Science Foundation under Grant
NSF PHY11-25915, the Netherlands Organization for Scientific Research
(NWO), the Marie Curie Initial Training Network `CosmoComp'
(PITN-GA-2009-238536), the European Research Council under the European Union’s Seventh Framework
Programme (FP7/2007-2013) via grant agreement 278594-GasAroundGalaxies and the Advanced Investigator grant COSMIWAY, and an STFC rolling grant awarded to the ICC. 


\bibliographystyle{mn2e}
\bibliography{bibliography} 
\bsp


\appendix
\medskip

\section{Numerical convergence}
\label{sec:numerical_convergence}

We next consider the effects of numerical resolution. Focussing on the \ZFe-\lx\ scaling relations, we follow the methodology adopted by C09, \citet{Font_et_al_11} and \citet{McCarthy_et_al_12a}, and compare the $Z_{\rm Fe}-L_{\rm X}$ relation derived from the intermediate-resolution simulations with results from the high-resolution realisation of the $-2\sigma$ \gimic\ region. The high resolution realisations have a factor of eight better mass resolution than the intermediate-resolution realisations (see \S~\ref{sec:simulations}).

\begin{figure*}
\includegraphics[width=\columnwidth]{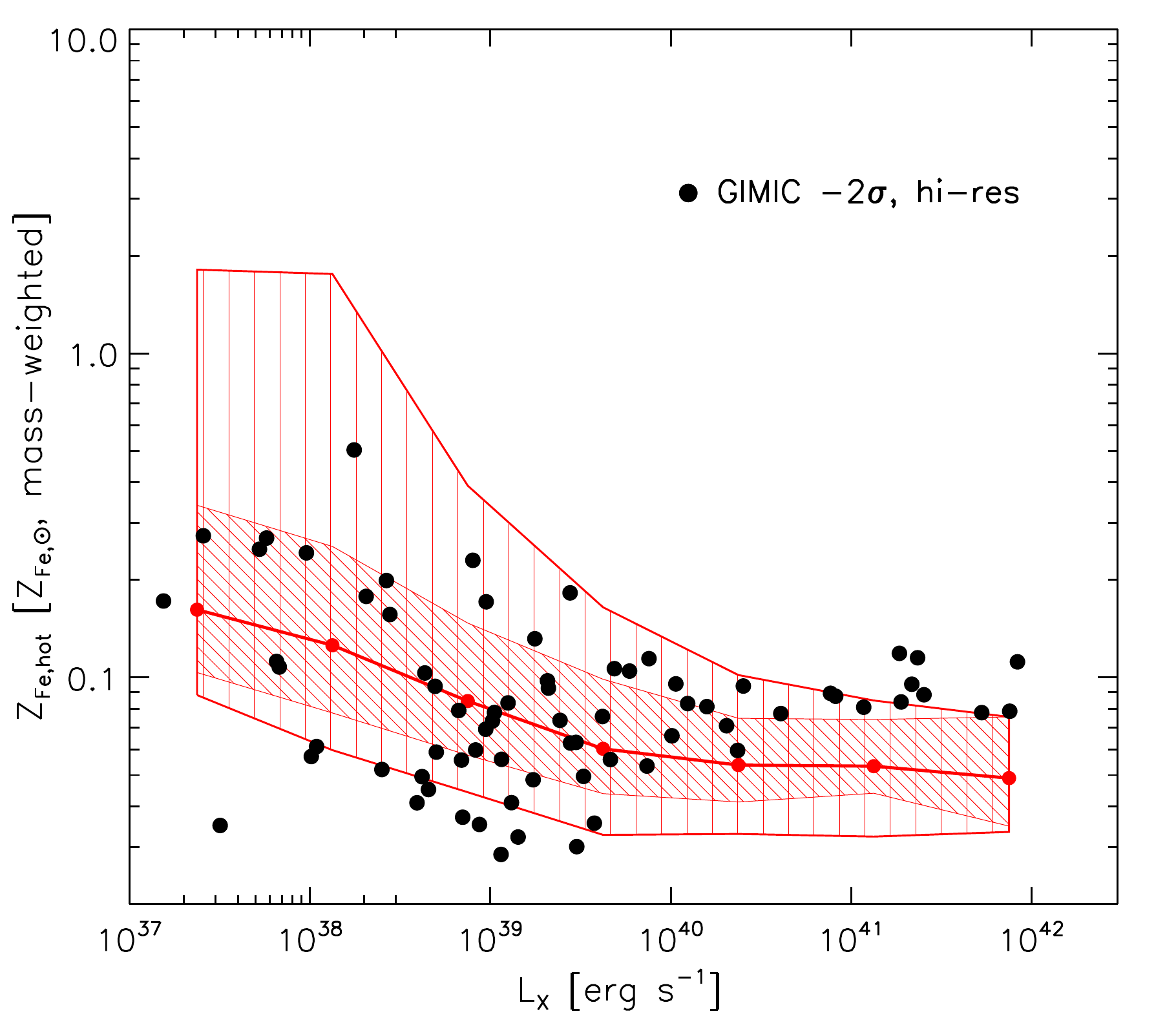}
\includegraphics[width=\columnwidth]{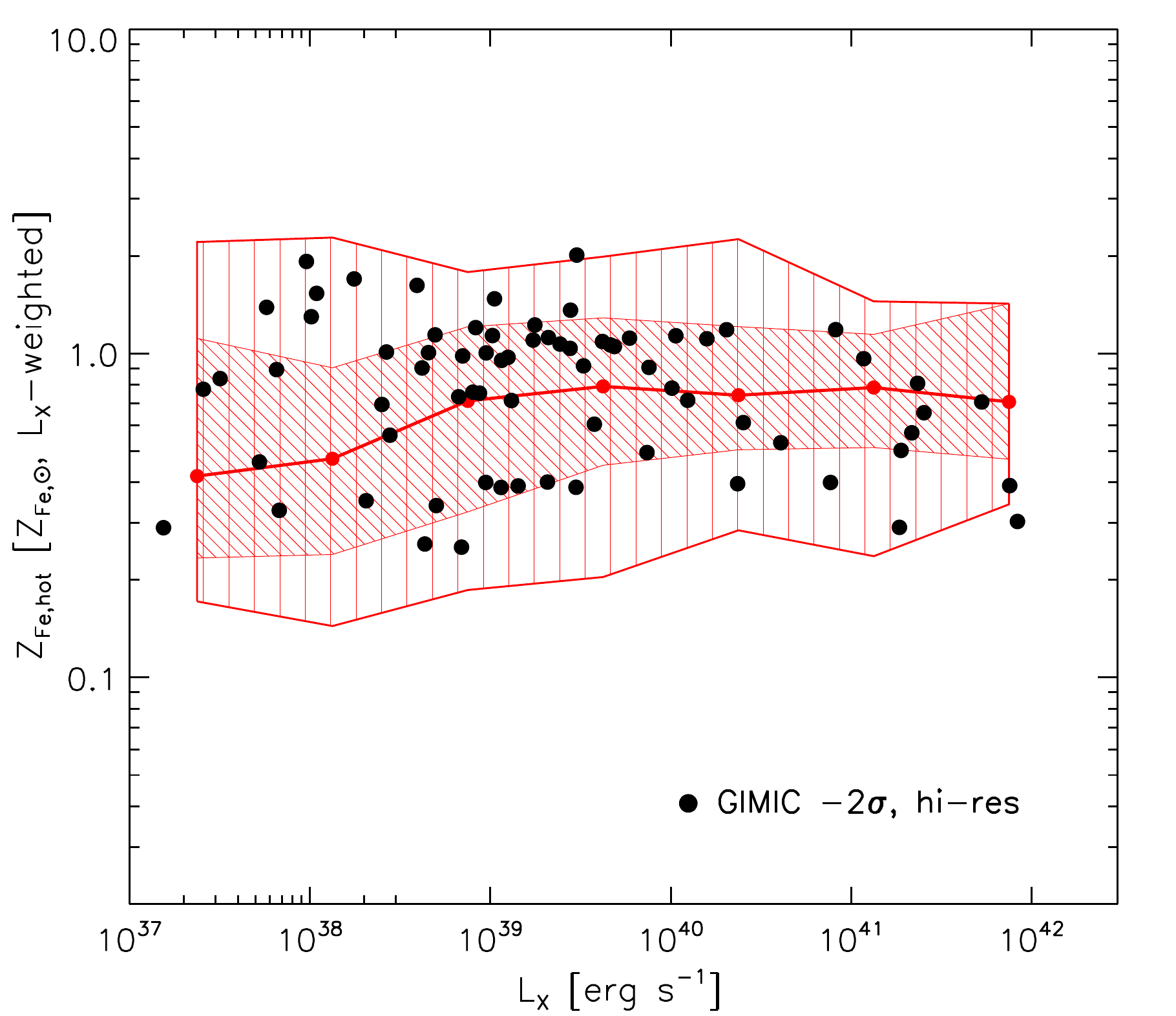}
\caption{The mass-weighted (\textit{left}) and luminosity-weighted  (\textit{right}) $Z_{\rm Fe}-L_{\rm X}$ relations defined by the 74 galaxies satisfying our selection criteria in the high-resolition $-2\sigma$ \gimic\ simulation, overplotted on the median and $(1,2)\sigma$-scatter (\textit{hatched regions}) defined by the sample of 617 galaxies drawn from all five intermediate-resolution simulations. The relations described by the intermediate- and high-resolution simulations are numerically converged.}
\label{fig:numerical_convergence}
\end{figure*}

In Fig.~\ref{fig:numerical_convergence}, we compare the mass-weighted
(\textit{left}) and \lx-weighted (\textit{right})
metallicity-luminosity scaling of the 74 galaxies satisfying our
selection criteria (see \S~\ref{sec:sample}) in the high-resolution
$-2\sigma$ region (\textit{black points}), with the median and scatter
derived from the sample of 617 galaxies in all five
intermediate-resolution regions (\textit{red hatching}). The
correspondence between the high- and intermediate-resolution samples
is good, indicating that the $Z_{\rm Fe}-L_{\rm X}$ relations (both
mass-weighted and \lx-weighted) are numerically convergerged in the
intermediate resolution simulations.

\section{Metallicity smoothing}
\label{sec:metallicity_smoothing}

In \S~\ref{sec:metal_mixing}, we discussed the lack of implicit scalar quantity diffusion in SPH, and the consequent suppression of metal mixing that unavoidably results from the finite sampling of the fluid. Here, we suppress the effects of suppressed mixing by kernel-weighted averaging element abundances over a given particles's SPH neighbours when computing cooling rates, radiative luminosities, and coronal metallicities. 

This smoothing procedure induces only a small error in the resulting total metal mass  \citep[see][]{Wiersma_et_al_09}\footnote{Strict conservation is not achieved because metals are distributed with a gather, rather than scatter, approach}, but does have the potential to bias the recovered metallicity of hot circumgalactic particles, because smoothed metallicities are computed using all SPH neighbour particles, irrespective of their thermal state. The smoothed element abundances of hot particles close to the disc-corona interface are therefore computed from both cold (i.e. ISM) and hot (coronal) particles. The maximum possible effect of such a potential bias can be determined by recomputing the metallicity and X-ray luminosity of each galaxy's hot CGM, using particle abundances instead of the SPH-smoothed abundances. 

\begin{figure*}
\includegraphics[width=\textwidth]{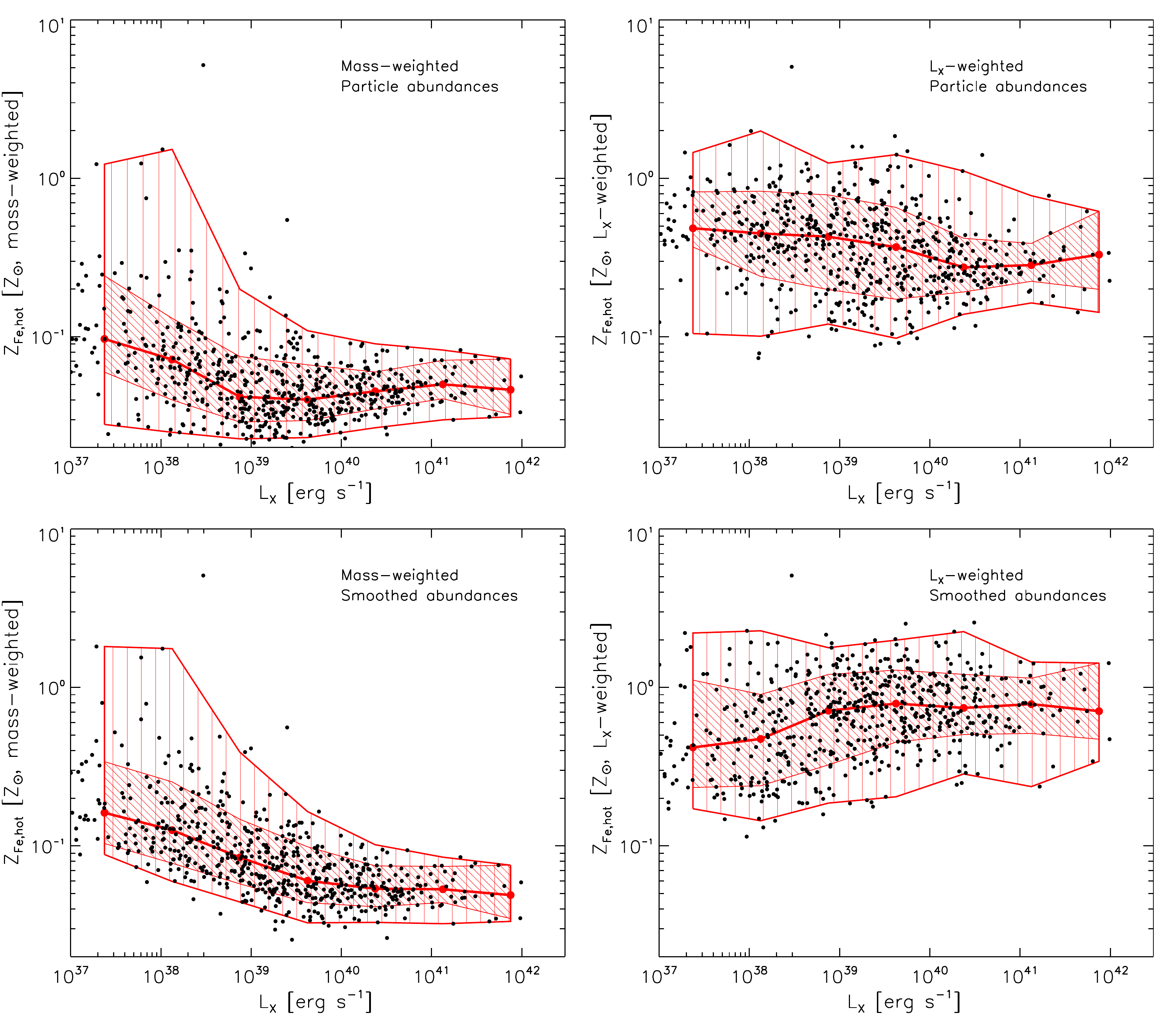}
\caption{Comparison of the mass-weighted (\textit{left-hand column}) and
  \lx-weighted (\textit{right-hand column}) $Z_{\rm Fe}-L_{\rm X}$
  relations recovered when adopting particle-based (\textit{top row})
  and smoothed (\textit{bottom row}) element abundances. The different
  abundance measures produce differences in the mass-weighted relation
  only for low-luminosity systems that are poorly resolved. The
  luminosity weighted relation exhibits differences for all systems
  probed here, but only at the factor $\lesssim 3$ level. Our
  conclusions are therefore insensitive to the use of the particle-based or smoothed element abundances when measuring coronal metallicities and computing particle X-ray luminosities.}
\label{fig:ZFe_Lx_particle_vs_smoothed}
\end{figure*}

In Fig.~\ref{fig:ZFe_Lx_particle_vs_smoothed}, we compare the mass-weighted (\textit{left-hand column}) and \lx-weighted (\textit{right-hand column}) $Z_{\rm Fe}-L_{\rm X}$ relations derived from the particle (\textit{top row}) and smoothed (\textit{bottom row}) abundances. As in previous figures, each dot represents a simulated galaxy, whilst the densely- and sparsely-hatched regions represent the $1\sigma$ and $2\sigma$ scatter, respectively. 

It is clear that the total X-ray luminosity is insensitive to metal mixing uncertainties, since no systematic shift in \lx\ is observed. Similarly, the recovered mass-weighted metallicity is little changed in well-resolved coronae, since the smoothing broadly conserves metal mass. In the lowest luminosity systems, which are less well resolved than more massive counterparts, the coronal metallicity derived from the particle abundances is lower by a factor of $\lesssim 2$, indicating that the degree of biasing caused by the smoothing of metals from cold ISM is small. It is, for example, smaller than the uncertainty in our adopted metal yields.

The effect on the luminosity-weighted metallicity is slightly larger: metallicities derived from particle abundances are lower by a factor of $\lesssim 3$. In contrast to the mass-weighted quantities, this result is relatively insensitive to the coronal luminosity, because the most luminous particles are always those closest to the disc-corona interface. Therefore, the extent of the largest possible bias resulting from our use of smoothed abundances is only marginally greater than the intrinsic uncertainty in our adopted metal yields, and does not impact upon our conclusions.

\section{Properties inferred by HB06 \& Ath07 from \chandra\ spectroscopy}
\label{sec:spectroscopy_comparison}

As described in \S~\ref{sec:observational_data}, we use the HB06
sample as our fiducial sample and supplement this with unique galaxies
from the Ath07 sample, since these extend to lower X-ray luminosity.
It is interesting briefly to examine the correspondence between the
values of \ZFe\ and \lx\ recovered by both studies from the archival
\chandra\ spectroscopy for the 22 galaxies that were included in both
samples, in order to assess the impact of systematic uncertainties.

\begin{figure*}
\includegraphics[width=\columnwidth]{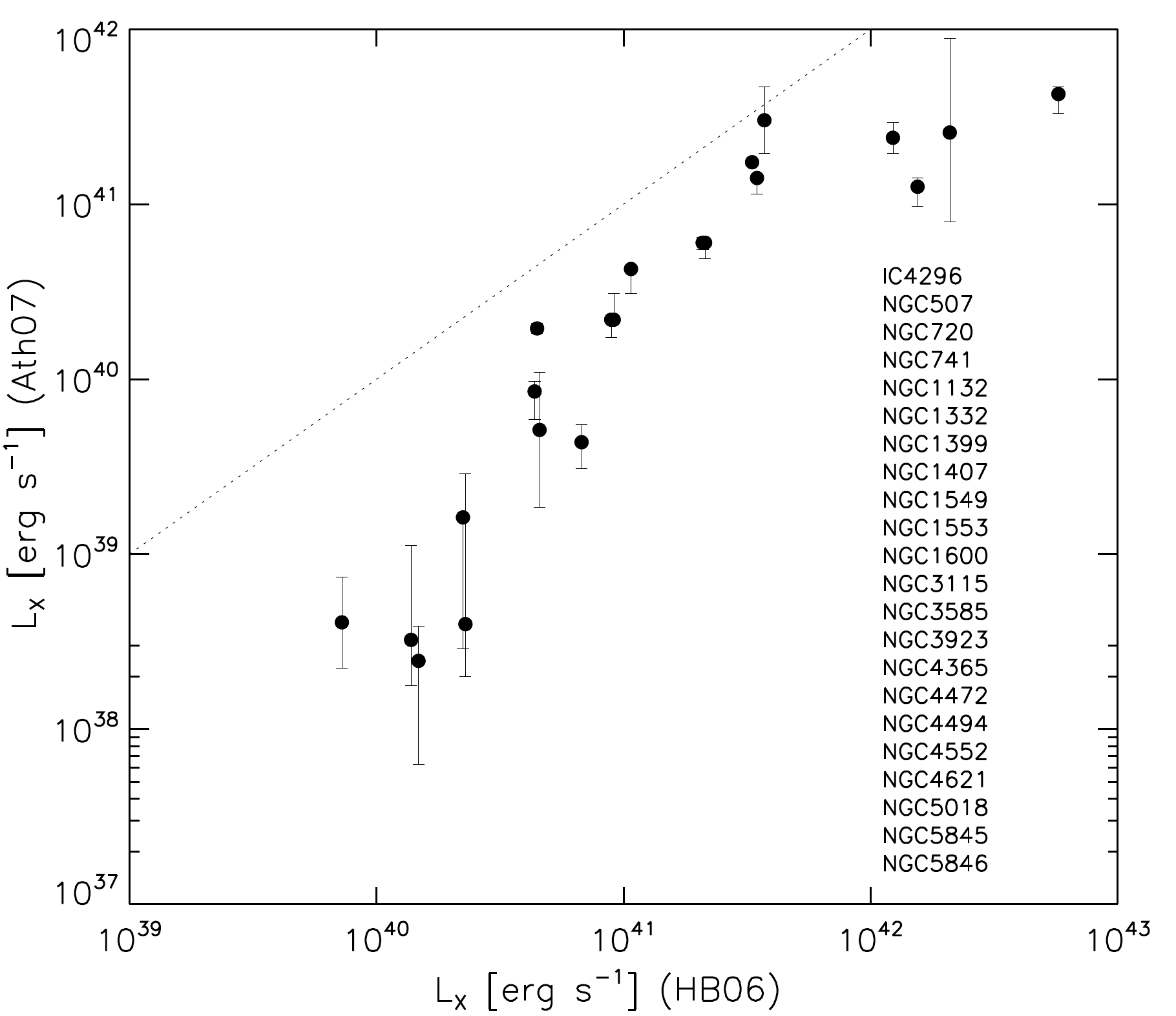}
\includegraphics[width=\columnwidth]{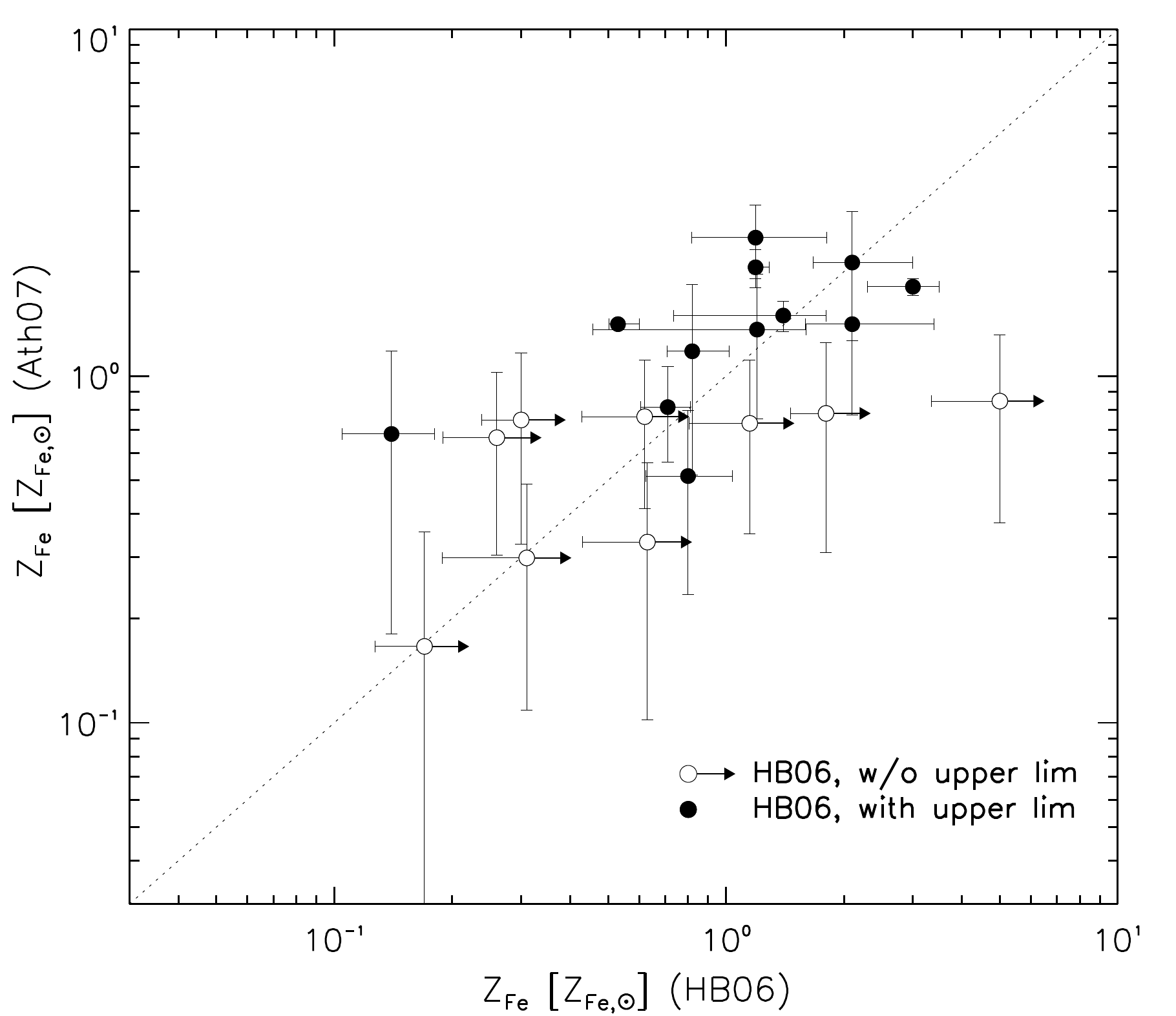}
\caption{The X-ray luminosity (\textit{left}) and iron abundance
  (\textit{right}) inferred by Ath07 as a function of that inferred by
  HB06, for the 22 galaxies present in both samples. The names of
  these galaxies are listed in the left-hand panel. Luminosity
  uncertainties are quoted only by Ath07. As in
  Fig.~\ref{fig:ZFe_Lx_Lk}, we plot galaxies from HB06 with no formal
  upper limit on \ZFe\ as open circles with arrows towards higher
  values. Iron abundances inferred by both studies are mostly
  consistent, and follow the 1:1 trend, albeit with significant
  scatter. Ath07 report systematically lower luminosities than HB06
  for the lowest luminosity galaxies. Our conclusions are, however,
  robust to these uncertainties.}
\label{fig:obs_convergence}
\end{figure*}

In Fig.~\ref{fig:obs_convergence}, we plot the soft X-ray luminosity
(\textit{left}) and the iron abundance (\textit{right}) of hot
circumglactic gas, reported by Ath07, as a function of that reported
by HB06. In each case, the 1:1 correspondence is shown as a dotted
line. Ath07 report systematically lower luminosities, with the
faintest and brightest galaxies differing by over a decade in \lx,
whilst intermediate luminosity galaxies (\lx$\sim10^{42}\ergs$) are
offset by $\sim0.5$dex. The inferred metallicities show no such
systematic offset, and broadly trace the 1:1 locus, albeit with broad
scatter.

The differences between the measurements highlight the difficulty of
reducing X-ray spectra, and fitting plasma models to these data to
infer luminosities and metallicities. However, the uncertainties
apparent here do not affect our conclusions: the inferred
metallicities are not systematically offset, whilst the application of
a systematic offset to the luminosties reported by either HB06 or
Ath07 would merely require the data to be translated along the x-axis
in the left-hand panel of Fig.~\ref{fig:ZFe_Lx_Lk}. Since the
simulated galaxies exhibit a flat \ZFe-\lx\ relation, such a
translation would leave the simulations entirely compatible with the
data.

\label{lastpage}
\end{document}